%% file: metbub.tex
\documentclass[useAMS,usenatbib]{mn2e}
\usepackage{amssymb,amsmath}
\usepackage{graphicx}
\usepackage{natbib}
\usepackage{journal_shortcuts}
\newcommand{\Sun}{_{\sun}}
\newcommand{\Rot}{_{\mathrm{rot}}}
\newcommand{\Max}{_{\mathrm{max}}}
\newcommand{\Min}{_{\mathrm{min}}}
\newcommand{\New}{_{\mathrm{new}}}
\newcommand{\Old}{_{\mathrm{old}}}
\newcommand{\Cluster}{_{\mathrm{cluster}}}
\newcommand{\Metal}{_{\mathrm{metal}}}
\newcommand{\Bubble}{_{\mathrm{bbl}}}
\newcommand{\Inj}{_{\mathrm{inj}}}
\newcommand{\Evac}{_{\mathrm{evac}}}
\newcommand{\ICM}{_{\mathrm{ICM}}}
\newcommand{\K}{\,\textrm{K}}
\newcommand{\Kpc}{\,\textrm{kpc}}
\newcommand{\Myr}{\,\textrm{Myr}}
\newcommand{\Gyr}{\,\textrm{Gyr}}
\newcommand{\Kms}{\,\textrm{km}\,\textrm{s}^{-1}}
\newcommand{\Erg}{\,\textrm{erg}}
\newcommand{\CCM}{\,\textrm{cm}^{-3}}
\newcommand{\gccm}{\,\textrm{g}\,\textrm{cm}^{-3}}
\newcommand{\ii}{\item}
\newcommand{\bi}{\begin{itemize}}
\newcommand{\ei}{\end{itemize}}

\title[Metal mixing by buoyant bubbles in galaxy clusters]%
{Metal mixing by buoyant bubbles in galaxy clusters}

\author[E. Roediger et al.]{E.~Roediger$^{1}$, M.~Br\"uggen$^{1}$, P.~Rebusco$^{2}$, H.~B\"ohringer$^{3}$, E.~Churazov$^{2,4}$ \\
$^1$ International University Bremen, P.O. Box 750\,561, 28725 Bremen,
Germany\\
$^2$ Max-Planck-Institut f\"ur Astrophysik, Karl-Schwarzschild-Strasse 1, 85741
Garching, Germany\\
$^3$ MPI f\"{u}r Extraterrestrische Physik, P.O. Box 1603, 85740
Garching, Germany\\
$^4$ Space Research Institute (IKI), Profsoyuznaya 84/32, Moscow 117810, 
Russia
}

\begin{document}

\date{Accepted. Received; in original form }

\pagerange{\pageref{firstpage}--\pageref{lastpage}} \pubyear{2005}

\maketitle

\label{firstpage}

\begin{abstract}

Using a series of three-dimensional, hydrodynamic simulations on an
adaptive grid, we have performed a systematic study on the effect of
bubble-induced motions on metallicity profiles in clusters of
galaxies. In particular, we have studied the dependence on the bubble
size and position, the recurrence times of the bubbles, the way these
bubbles are inflated and the underlying cluster profile.  We find that
in hydrostatic cluster models, the resulting metal distribution is
very elongated along the direction of the bubbles. Anisotropies in the
cluster or ambient motions are needed if the metal distribution is
to be spherical. In order to parametrise the metal transport by
bubbles, we compute effective diffusion coefficients. The diffusion
coefficients inferred from our simple experiments lie at values of
around $\sim 10^{29}$ cm$^2$s$^{-1}$ at a radius of 10 kpc. The runs
modelled on the Perseus cluster yield diffusion coefficients that
agree very well with those inferred from observations.

\end{abstract}

\begin{keywords}

\end{keywords}

%
%
%
%
\input{intro}

\input{method}

\input{results}
\input{diff_const}
\input{discussion}

\input{summary}


\section*{Acknowledgements}

We thank Mateusz Ruszkowski and Matthias Hoeft for helpful
discussions. The anonymous referee has made a number of very
useful suggestions to improve the paper. Furthermore, we acknowledge
the support by the DFG grant BR 2026/3 within the Priority Programme
``Witnesses of Cosmic History'' and the supercomputing grants NIC 1927
and 1658 at the John-Neumann Institut at the Forschungszentrum
J\"ulich.  Some of the simulations were produced with STELLA, the
LOFAR BlueGene/L System in Groningen.

The results presented were produced using the FLASH code, a product of the DOE
ASC/Alliances-funded Center for Astrophysical Thermonuclear Flashes at the
University of Chicago.


%
\bibliographystyle{mn2e}
\bibliography{%
../BIBLIOGRAPHY/radio,%
../BIBLIOGRAPHY/metals%
}

\bsp

\label{lastpage}

\end{document}

%% file: intro.tex
\section{Introduction}
%

The hot, diffuse gas that permeates clusters of galaxies, the
intra-cluster medium (ICM), has a metallicity of about 1/3 of the
solar value. Meanwhile, the total amount of iron in the ICM is huge:
it is larger than the total iron mass in all cluster
galaxies. Moreover, this value does not seem to vary much with time,
at least not up to a redshift of $\sim 1$ (\citealt{tozzi:03}). The
recent generation of X-ray observatories has provided maps of the
radial distribution of metals in the nearby clusters
(\citealt{degrandi:01,fukazawa:00,schmidt:02,matsushita:02,churazov:03,degrandi:04,tamura:04}). These
observations have revealed an interesting trend:

Galaxy clusters can be grouped into two categories depending on their
X-ray surface brightness profiles: (i) clusters with a central peak in
the X-ray surface brightness, i.e. clusters with a cool core and (ii)
clusters without a cool core. Interestingly, these two groups show a
different spatial distribution of metals. The clusters without cool
cores have a nearly uniform spatial distribution of metals, while
clusters with cool cores have a strongly peaked abundance
profile. Moreover, the relative abundance of elements in cool core
clusters suggests that Type Ia supernovae have provided most of the
metals in the central iron peak. \cite{finoguenov:02} have found that
the central abundance peak is strongly enriched in iron while the bulk
of the ICM outside the central region had an iron-to-silicon ratio
close to the yields of Type II supernovae (also
\citealt{mushotzky:96}). This suggests that the central metal peak is
mostly caused by Type Ia supernovae in the central cluster
galaxy. However, in order to produce the large mass of metals in the
central peak, long enrichment times are necessary.

There are a number of observations that indicate that stars in massive
galaxies are responsible for the metal enrichment of the ICM.  It has
been established that the iron mass of the ICM correlates with the
optical light of massive early-type galaxies in clusters
(\citealt{arnaud:92}) and that the ratio of the iron mass to light and
to the total mass is approximately constant. Thus, stellar mass loss
from the central cluster galaxy through supernovae and stellar winds
appears to be the prime source for the metals observed in the inner
parts of galaxy clusters. However, the observed metallicity profiles
are much broader than the stellar light profiles of the central
galaxy. Hence, the differences in the light and metal distributions
are interpreted as the result of transport processes that have mixed
the metals into the ICM. \cite{rebusco:05} have assumed that the
spreading is due to local turbulent motions in the ICM, and that the
evolution of the metal density profile can be modelled as a diffusion
process.  Using the abundance profile from \cite{churazov:03} for
Perseus, \cite{rebusco:05} estimated that a diffusion coefficient of
the order of $2\times10^{29}~{\rm cm^2~s^{-1}}$ is needed to explain
the width of the observed abundance profiles. Based on a model of iron
enrichment from stellar mass loss in the central cluster galaxies of a
few nearby clusters, \cite{boehringer:04} have used the total amount
of iron observed in the ICM to obtain constraints on the age of the
cooling cores (i.e. the time for which the central region have
remained relatively undisturbed). They find very large ages of more
than 7 Gyr (\citealt{boehringer:04}) that provide new constraints for
the modelling of the interaction regions and have consequences for the
turbulent transport of metals. While it appears to be established that
the metals produced by the central galaxy are dispersed into the ICM
to form the broad abundance peaks, it remains unclear what the
mechanism is via which the metals are transported.\\

Currently, the most popular model that is invoked to explain the
apparent stability of cool cores against a cooling catastrophe relies
on heating by a central Active Galactic Nucleus (AGN). Radio-loud AGN
are also thought to regulate the formation of the most massive
galaxies and thus explain the cut-off of the galaxy luminosity
function at the bright end (\citealt{croton:06}). Radio-loud active
galactic nuclei drive strong outflows in the form of jets that inflate
bubbles or lobes. The lobes are filled with hot plasma, and can heat
the cluster gas in a number of ways (e.g. \citealt{bruggen:02,
bruggen:02a,churazov:01,dallavecchia:04,omma:04,basson:03,reynolds:02,ruszkowski:04,bruggen:05,brighenti:06}).\\

In fact, radio sources in cooling clusters are different from the
general population of radio sources in galaxy clusters. Most radio
galaxies in clusters are FR~I sources in terms of their radio power,
but they show a great variety of morphologies and dynamics
(\citealt{owen:97}).  The central galaxy in almost every strong
cooling core contains an active nucleus and a currently active,
jet--driven radio galaxy. High-resolution X-ray observations of cooling flow clusters
with {\sc Chandra} have revealed a multitude of X-ray holes on scales
$<$ 50 kpc, often coincident with patches of radio emission. In a
recent compilation, \cite{birzan:04} lists 18 well documented clusters
which show X-ray cavities with radio emission.\\


Buoyantly rising bubbles induce subsonic motions in the cluster gas
that can redistribute metals in the ICM
(\citealt{bruggen:02b}). Considering the impact of resonant
scatterings on the surface brightness profile of the most prominent
X-ray emission lines, evidence for gas motions was found in the
XMM-Newton data on the Perseus cluster (\citealt{churazov:04}). The
same motions could spread the metals over the cluster volume although
the efficiency of spreading could be very sensitive to the character
of these motions. Other enrichment mechanisms include
supernovae-driven winds, radiation-pressure-driven dust efflux (see
e.g. \citealt{aguirre:01}) and ram pressure stripping of cluster
galaxies (\citealt{roediger:05, roediger:06}). Using cosmological
simulations and a heuristic prescription for the effect of
ram-pressure stripping,
\cite{domainko:06} find that ram-pressure stripping can account for
$\sim$10\% of the overall observed level of enrichment in the ICM
within a radius of 1.3~Mpc (see also
\citealt{tornatore:04}). Furthermore,
\cite{schindler:05} find that ram-pressure stripping is more efficient
than quiet (i.e. non-starburst driven) galactic winds, at least in
recent epochs since redshift 1. Also, they find that the expelled
metals are not mixed immediately with the ICM, but inhomogeneities are
visible in the metallicity maps. Relatively little theoretical work
has gone in establishing the importance of AGN-induced transport of
metals in cluster centres
(\citealt{bruggen:02b}, \citealt{omma:04}). However, this mode is
likely to be very important, at least at redshifts $\leq 1$: If
AGN-induced motions are sufficient to quench cooling flows at the
centres of galaxy clusters, they are similarly likely to affect the
metal distribution in the central cluster region.  Simulations as
those presented here are essential for a proper modelling of the
chemical evolution of clusters via semi-analytical techniques, as for
example in
\cite{delucia:04,cora:06}.\\

The broad abundance peaks in clusters have been produced over a
time span of several gigayears. If the peaks have been broadened by
AGN, this process has taken a large number of activity cycles. In
order to simulate the transport of metals by AGN-induced flows and to
capture the hydrodynamical details important for the interaction
between the AGN and the ICM, a fair resolution of the computational
mesh is necessary. This prevents us from simulating the
bubble transport over cosmological times. However, we can still study
the efficiency of the metal transport by simulating a small number of
AGN cycles and parametrise the transport efficiency. A convenient
parametrisation that has been applied previously to observations
(\citealt{rebusco:05}) is based on a diffusion description. Even
though we do not propose that the metal transport occurs via
microscopic diffusion, transport processes like these may be described
by a diffusion parameter. This allows a direct comparison to
observations without having to simulate the entire life time of the cluster.\\

In this paper, we investigate the influence of buoyantly rising
bubbles in clusters on the metal distribution in the ICM. In
particular, we focus on the following questions:

\begin{itemize}

\item How far out can bubbles carry metals?
\item How efficient is transport by buoyant bubbles?
\item Does this process cause characteristic features, e.g. anisotropies? 
\item How does the resulting metal distribution depend on bubble and cluster parameters?
\item What effective diffusion coefficients does this form of
transport yield?
\item How do these coefficients compare with those found by
\cite{rebusco:05}?

\end{itemize}

To this end, we simulated the effect of bubble-induced motions in the
ICM based on the 3D hydrodynamics code {\sc FLASH}.

%% file: method.tex
\section{Method}
%
\subsection{Overview}
We model the hydrodynamical evolution of the ICM in a galaxy
cluster. Initially, the ICM is set in hydrostatic equilibrium in a
static cluster potential. Details of the ICM model are given in
Sect.~\ref{sec:ICM}.

We assume that a central cluster galaxy injects metals into the ICM
with a rate proportional to its light distribution. Details of the
central galaxy are given in Sect.~\ref{sec:centr_gal}. We now trace
the distribution of the metals by injecting a tracer fluid into the
ICM at a rate that is proportional to the light distribution of the
central galaxy. Section~\ref{sec:method_metals} explains the details
of the metal injection and the way the metals are traced throughout
the simulation.

Finally, we model the AGN activity by inflating ambipolar pairs of
underdense bubbles in the ICM that rise buoyantly and thus stir the
ICM. In Section~\ref{sec:bubble_generation} we explain in detail how
the bubbles are generated.

\subsection{Code} \label{sec:code}
The simulations were performed with the FLASH code
(\citealt{fryxell:00}), a multidimensional adaptive mesh refinement
code. FLASH is a modular block-structured AMR code, parallelised using
the Message Passing Interface (MPI) library. It solves the Riemann
problem on a Cartesian grid using the Piecewise-Parabolic Method
(PPM). The simulations are performed in 3D and all boundaries are
reflecting. We use a simulation box of size $(x\Min,x\Max)\times
(y\Min,y\Max)\times (z\Min,z\Max)=(-145\Kpc,145\Kpc)^3$. The cluster
centre is located at $(x\Cluster,y\Cluster,z\Cluster)=(0,0,0)$. For
our grid, we chose a block size of $16^3$ zones. The unrefined root
grid contains $6^3$ blocks. The refinement criteria are the standard
density and pressure criteria. In the central $\sim 30\Kpc$, the grid
is assigned a minimum refinement level of 3. Outside $\sim 100\Kpc$, the refinement
is restricted to a maximum of 5 levels to limit the computational
effort. The resulting resolutions for the different runs are
summarised in Table~\ref{tab:runlist}.\\

There have been some suggestions that the ICM may have a
non-negligible viscosity
(\citealt{reynolds:05,ruszkowski:04}). However, the effective
viscosity of the ICM remains unknown because the physics of such
dilute and magnetised plasmas is poorly constrained. In particular,
the effect of magnetic fields on the macroscopic viscosity is
unclear. Even minute magnetic fields lead to small proton gyroradii
and are thus expected to suppress the viscosity efficiently. However,
it has been pointed out that the exponential divergence of neigbouring
field lines in a tangled magnetic field may lead to only a very modest
suppression of the viscosity.
\citet{reynolds:05} have simulated the buoyant rise of bubbles in a
medium with a kinematic viscosity. They found that even a
modest shear viscosity (corresponding to 1/4 of the Spitzer value) can
quench fluid instabilities and keep the bubbles intact (especially see
figure 4 in their paper). Other simulations of buoyant bubbles in a
viscous ICM were performed by \citet{sijacki:06} using smoothed
particle hydrodynamics.

FLASH solves the equations of inviscid hydrodynamics. The
Piecewise-Parabolic Method uses a monotonicity constraint rather than
artificial viscosity to control oscillations near
discontinuities. However, the discretisation of the equations still
leads to a numerical viscosity. 
A comparison with the simulations of bubbles in a viscous medium by
\citet{reynolds:05} (cf. with their figure 4) shows that the Reynolds
number in our simulations is $> 1000$.\\

Except by the effect of viscosity, bubbles can be stabilised by
magnetic fields. It has been known for some time that the hot gas in clusters of
galaxies hosts significant magnetic fields \citep{carilli:02}, with
estimated typical magnetic field strengths of $5 \times 10^{-6}$ G (5
$\mu$G).  A magnetic field tangential to the fluid interface can
stabilise the Rayleigh-Taylor instability \citep{chandrasekhar:61},
provided the restoring tension generated by bending of the field lines
exceeds the buoyancy force driving the instability. The stabilising
effect of intracluster magnetic fields was pointed out by
\cite{deyoung:03}, who derived analytic conditions for the
stabilisation of relic radio bubbles in the hot ICM. These calculations
showed that the field strengths observed in many clusters will
stabilise the bubble interface in the linear regime, and hence the
cluster magnetic fields could account for why the relic radio bubbles
are seen as intact objects at such late times. Additional support for
this idea is found in the two-dimensional MHD calculations of
\cite{bruggen:01}, who applied a $\beta \sim 10$ field inside the
bubbles that was aligned with the bubble surface. While viscosity and
magnetic fields can affect the dynamics of the bubbles and thus also
the resultant transport of metals, the effects are still poorly
understood. Here we only study the case of inviscid and unmagnetised fluids.

\subsection{Model cluster}

\subsubsection{ICM distribution} \label{sec:ICM}

In this paper, we study two different cluster models. In the first,
generic one, we take a constant ICM temperature and the
pressure is set by assuming hydrostatic
equilibrium. Table~\ref{tab:ICM} lists the ICM parameters.

The density of the ICM follows a $\beta$-profile, 
%
\begin{equation}
\rho\ICM (r) = \rho\ICM{}_0 \left[ 1+\left( \frac{r}{R\ICM}  \right)^2  \right]^{-3/2\beta} 
\end{equation}
%

%
\begin{table}
\caption{ICM parameters for generic runs.}
\label{tab:ICM}
\centering\begin{tabular}{ll}
\hline
$R\ICM$        &  $50\Kpc$   \\
$\rho\ICM{}_0$ &  $10^{-26}\gccm$ \\
$\beta$        &  $0.5$  \\
$T\ICM$        &  $4.7\cdot 10^7\K$ \\
\hline
\end{tabular}
\end{table}

In addition to the generic cluster model, we set up a cluster
modelled on the brightest X-ray cluster A426 (Perseus) that has been studied
extensively with {\sc Chandra} and {\sc XMM}-Newton.

The electron density $n_{\rm e}$ and the temperature $T_{\rm e}$ profiles used here are
based on the deprojected XMM-Newton data (Churazov et al. 2003, 2004)
which are also in broad agreement with the ASCA (Allen \& Fabian
1998), Beppo-Sax (De Grandi \& Molendi 2001, 2002) and Chandra
(Schmidt et al. 2002, Sanders et al. 2004) data. Namely:

\begin{eqnarray}
n_{\rm e}=\frac{4.6\times10^{-2}}{[1+(\frac{r}{57})^2]^{1.8}}+
\frac{4.8\times10^{-3}}{[1+(\frac{r}{200})^2]^{0.87}}~~~{\rm cm}^{-3}
\label{ne}
\end{eqnarray}

and

\begin{eqnarray}
T_{\rm e}=7\times\frac{[1+(\frac{r}{71})^3]}{[2.3+(\frac{r}{71})^3]}~~~{\rm keV},
\label{te}
\end{eqnarray}

where $r$ is measured in kpc. The hydrogen number density is assumed
to be related to the electron number density as $n_{H}=n_{\rm e}/1.2$.
%


\subsubsection{Central galaxy} \label{sec:centr_gal}

The metal injection rate in the central galaxy is assumed to be
proportional to the light distribution.  It is modelled with a
Hernquist profile given by:

\begin{equation}
\dot\rho\Metal (r) = \frac{\dot M\Metal{}_0}{2\pi} \, \frac{a}{r} \, \frac{1}{(r+a)^3}
\label{eq:hernquist}
\end{equation}
%
Note that the injection rate could also be
time-dependent. \cite{rebusco:05} have used a time-dependent metal
injection rate that accounts for the higher supernova rate in the past
and the evolution of the stellar population (also
see\citealt{renzini:93}).  However, we follow the evolution of the
cluster for only about $1\Gyr$. For typical cases studied in
\cite{rebusco:05}, the metal injection rate does not change
significantly over this time. Hence in our simulations we use a
temporally constant metal injection rate. Table~\ref{tab:metinj} lists
the parameters for the metal injection.
%
\begin{table}
\caption{Central galaxy -- metal injection parameters.}
\label{tab:metinj}
\centering\begin{tabular}{ll}
\hline
$a$                &  $10\Kpc$   \\
$\dot M\Metal{}_0$ &  $952\,M\Sun\Myr^{-1}$ \\
\hline
\end{tabular}
\end{table}
%

\subsection{Tracing the metals} \label{sec:method_metals}
The FLASH code offers the opportunity to advect mass scalars along with the
gas density.  In order to be able to trace the metal distribution, we utilise
one of the mass scalars, $f$, to represent the local metal fraction in each
cell:
%
\begin{equation}
f=\frac{\rho\Metal}{\rho}
\end{equation}
%
Hence, the quantity $f\rho$ gives the local metal density, $\rho\Metal$. The
metal density obeys the continuity equation including the metal source:
%
\begin{equation}
\frac{\partial}{\partial t}  (f\rho)  = - \nabla\cdot (f\rho \,\vec v) +
\dot\rho\Metal. \label{eq:continuity_metals}
\end{equation}
%
In order to model the metal injection by the central galaxy, we assume that
the metal fraction is small at all times. Hence, we can neglect
$\dot\rho\Metal$ as a source term in the continuity equation for the gas
density, $\rho$. As a consequence, the system of hydrodynamical equations
including Eq.~\ref{eq:continuity_metals} is identical for the set of variables
$(\dot\rho\Metal,f)$ and $(A\cdot\dot\rho\Metal,A\cdot f)$, where $A$ is a
constant. In other words, the amplitude of the metal injection is arbitrary,
$f\rho$ and $f$ can be scaled to the total metal injection rate of the central
galaxy, $\dot M\Metal{}_0$. In the numerical implementation, the metal
fraction at each position $\vec x$ is updated according to
%
\begin{equation}
f\New(\vec x) = \frac{f\Old(\vec x)\rho(\vec x) + \dot\rho\Metal(\vec
  x)\cdot \Delta t}{\rho(\vec x)}
\end{equation}
%
in each timestep.

\subsection{Bubble generation} \label{sec:bubble_generation}

Bubbles in the ICM are thought to be inflated by a pair of ambipolar
jets from an AGN in the central galaxy. In this simple picture, the
jets inject energy into small volumina of the ICM at their terminal
points. Thus, this region is in overpressure and expands until it
reaches pressure equilibrium with the surrounding ICM. The result is a
pair of underdense, hot bubbles.

In our simulations, we generate pairs of bubbles on opposite
sides of the cluster centre. There are various ways in which one could
produce these bubbles and we have tested the following two methods:

\begin{itemize}

\item In the first method, we inflate the bubbles by injecting energy
  into two small spheres of radius $r\Inj$ with a constant rate $\dot
  e$ (in erg$\CCM\,$s) over an interval of length $\tau\Inj$. The
  total injected energy is $E\Inj=\dot e 4 \pi r\Inj^3 \tau\Inj$ per
  bubble. The gas inside these spheres is heated and expands similar
  to a Sedov explosion in a few Myr to form a pair of bubbles. The
  parameters $r\Inj$, $E\Inj$ and $\tau\Inj$ are chosen such that the
  resulting bubbles show approximately a density contrast of
  $\rho_{\rm b}/\rho_{\rm amb} = 0.03$ to the surrounding ICM and
  approximately a radius of $r\Bubble$. The expansion is much shorter
  than the rise of the generated bubbles. In addition to the bubbles,
  the explosion sets off shock waves that move through the
  ICM. Incidentally, the dependence of the bubble dynamics on the
  density contrast, $\rho_{\rm b}/\rho_{\rm amb}$, is weak provided
  that $\rho_{\rm b}/\rho_{\rm amb} \ll 1$.

\item In the second method, we evacuate the bubble regions by removing gas while keeping them in pressure equilibrium with their surroundings. This evacuation method does not produce the shock waves generated by the
  inflation method. We implement gas mass sinks inside two spheres of
  radius $r\Bubble$. Inside these regions, gas is removed with a
  certain rate $\dot\rho$ over a time $\tau\Evac$. The evacuation is
  not done instantaneously to prevent numerical problems. However,
  $\tau\Evac$ is small compared to the evolution timescale of the
  bubbles. The sink rate $\dot\rho$ is set to decrease the the density
  inside the bubbles down to a density contrast of $\rho_{\rm
  b}/\rho_{\rm amb}$ compared to the surrounding ICM. In order to
  conserve the metal mass, the metal fraction also needs to be updated
  during the evacuation.  The evacuation method does not cause shocks
  that move through the ICM and is, hence, computationally
  cheaper. The evacuation method models the phase of the
  evolution after the radio-loud AGN has inflated a low-density bubble
  which has expanded to achieve pressure equilibrium with its
  surroundings \citep{bruggen:01}. The inflation of the lobes itself
  cannot be modelled entirely self-consistently as one would have to
  resolve scales all the way down to the accretion disk of the
  AGN. However, jets are observed to inflate spherical bubbles that
  are in near pressure equilibrium with the ICM. In order to model the
  buoyant rise of these AGN-blown bubbles, it is customary to
  start with an underdense, spherical bubble. Since gas is removed
  from a small volume in the process of forming the bubble, mass is not
  conserved in this method. However, this does not affect the density
  profile of the cluster on the time scales that we consider here. In
  Fig. \ref{fig:prof_Mmets_evac_infl} we show that the differences
  between methods EVAC and INFL are very small.

\end{itemize}

For both methods, we generate a pair of bubbles every
$\tau\Bubble$. For every bubble generation cycle, the position of the
generated bubbles changes. For the next pair of bubbles, the axis that
connected the previous bubbles is rotated around the $y$-axis. The
angle of rotation is fixed by the demand that after $n\Rot{}\Bubble$
cycles the bubbles are again inflated in the first position. The
parameters for our runs are summarised in Table \ref{tab:runlist}.

%% file: results.tex
\section{Results}
%
Table~\ref{tab:runlist} gives an overview of the simulations described
in this paper. Primarily, we test the dependence of the resulting
metal distribution on the bubble position, size and recurrence time, the
background model, the method with which bubbles are produced and the
resolution of the computational grid.

\begin{table*}
\caption{Simulation runs. ``$\cdot$'' means same value as in previous
  line. ``/'' means value irrelevant for this run. The first column gives the coordinates of the initial bubbles centres. $r\Bubble$ is the radius of the bubble, $\tau\Bubble$ is the recurrence time of the bubbles, $E\Inj$ the total injected energy, $n\Rot{}\Bubble$ the number of precessions of the bubble, $(x/y/z\Min{}_/{}\Max)$ the size of the computational domain, and $\Delta x$ the side length of the smallest computational cell.}
\label{tab:runlist}
\centering\begin{tabular}{lcccccccc}
\hline
name & $(x\Bubble, y\Bubble, z\Bubble)$ & $r\Bubble$ &
$\tau\Bubble$ & $E\Inj$ &
$n\Rot{}\Bubble$ & $(x/y/z\Min{}_/{}\Max)$ & effective & $\Delta x$ \\
     & in $(\Kpc,\Kpc,\Kpc)$ & $/\Kpc$ &  &
$/\Myr$       & $/\Erg$ &
              & no of cells & (kpc)  \\
\hline
SINGLE\_BBL & (10,12,0) &    10   &   /     &
      /              & / & $\pm146$      & $768^3$ & $0.38$ \\
\hline
EVAC           & $\cdot$   & $\cdot$ &    50   &
      /              &  4  & $\cdot$ & $192^3$ &  $1.52$ \\
EVAC\_HR       & $\cdot$   & $\cdot$ & $\cdot$ & 
      /              & $\cdot$ & $\cdot$ & $384^3$ & $0.76$ \\
\hline
INFL           & $\cdot$   & $\cdot$ & $\cdot$ &
$6.68\cdot 10^{57}$  & $\cdot$ & $\cdot$ & $192^3$ &  $1.52$ \\
INFL\_HR       & $\cdot$   & $\cdot$ & $\cdot$ &
    $\cdot$          & $\cdot$ & $\cdot$ & $384^3$ & $0.76$ \\
\hline
TAUBBL200      & $\cdot$ & $\cdot$ &   200   &
     evac      & $\cdot$ & $\pm340$ & $448^3$ & $1.52$ \\
\hline
DISTCTR        & (20,24,0) & $\cdot$ & 50 &
     $\cdot$   & $\cdot$ & $\cdot$ & $\cdot$ & $\cdot$ \\
LARGEBBL       & $\cdot$ & 20 & $\cdot$ &
     $\cdot$   & $\cdot$ & $\cdot$ & $\cdot$ & $\cdot$ \\
\hline
PERSEUS\_50 & (10,12,0) &    10   &  50     &
      /              &  4  & $\pm146$ & $192^3$ &  $1.52$ \\
PERSEUS\_200 & (10,12,0) &    10   &  200     &
      /              &  4  & $\pm146$ & $192^3$ &  $1.52$ \\

\hline
\end{tabular}
\end{table*}

\subsection{Stability test}

\begin{figure}
\centering\resizebox{\hsize}{!}%
{\includegraphics[angle=-90]{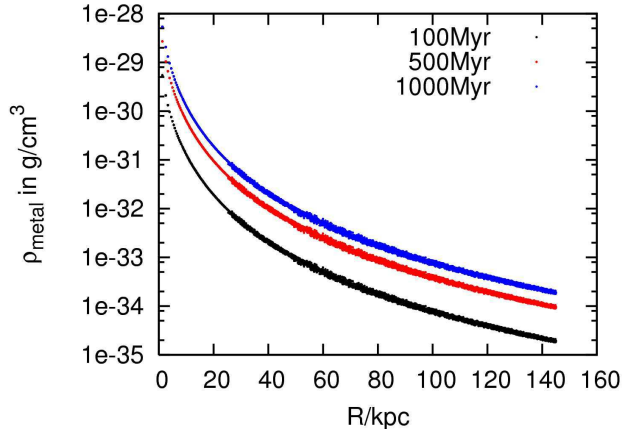}}
\caption{Evolution of the metal density profile in the case without bubbles. The metal density increases with time proportional to the light profile of the central galaxy. The injection rate is assumed to be constant in time. It is apparent that numerical mixing is very small. The noise increases at larger radii as the refinement levels go down.}
\label{fig:stabtest_metals}
\end{figure}

Before stirring the ICM with an AGN, we tested whether the cluster
remains in hydrostatic equilibrium for a typical simulation time of
about 1 Gyr. To this end, we ran a simulation where we injected metals
but did not produce any bubbles. We found that, both, gas density and
metal density profiles remained stable for at least
$1\Gyr$. Figure~\ref{fig:stabtest_metals} shows the evolution of the
metal density profile.

Even after $1\Gyr$ the 3 refinement levels can be seen in the
profiles (where the ``line'' becomes thicker). Numerical velocities are of the
order of $1\Kms$ inside the grid and of $15\Kms$ at the grid boundary.

\subsection{Single pair of bubbles} \label{sec:singlebubble}

\begin{figure*}
\includegraphics[trim=0 80 0 0,clip,width=0.45\textwidth]{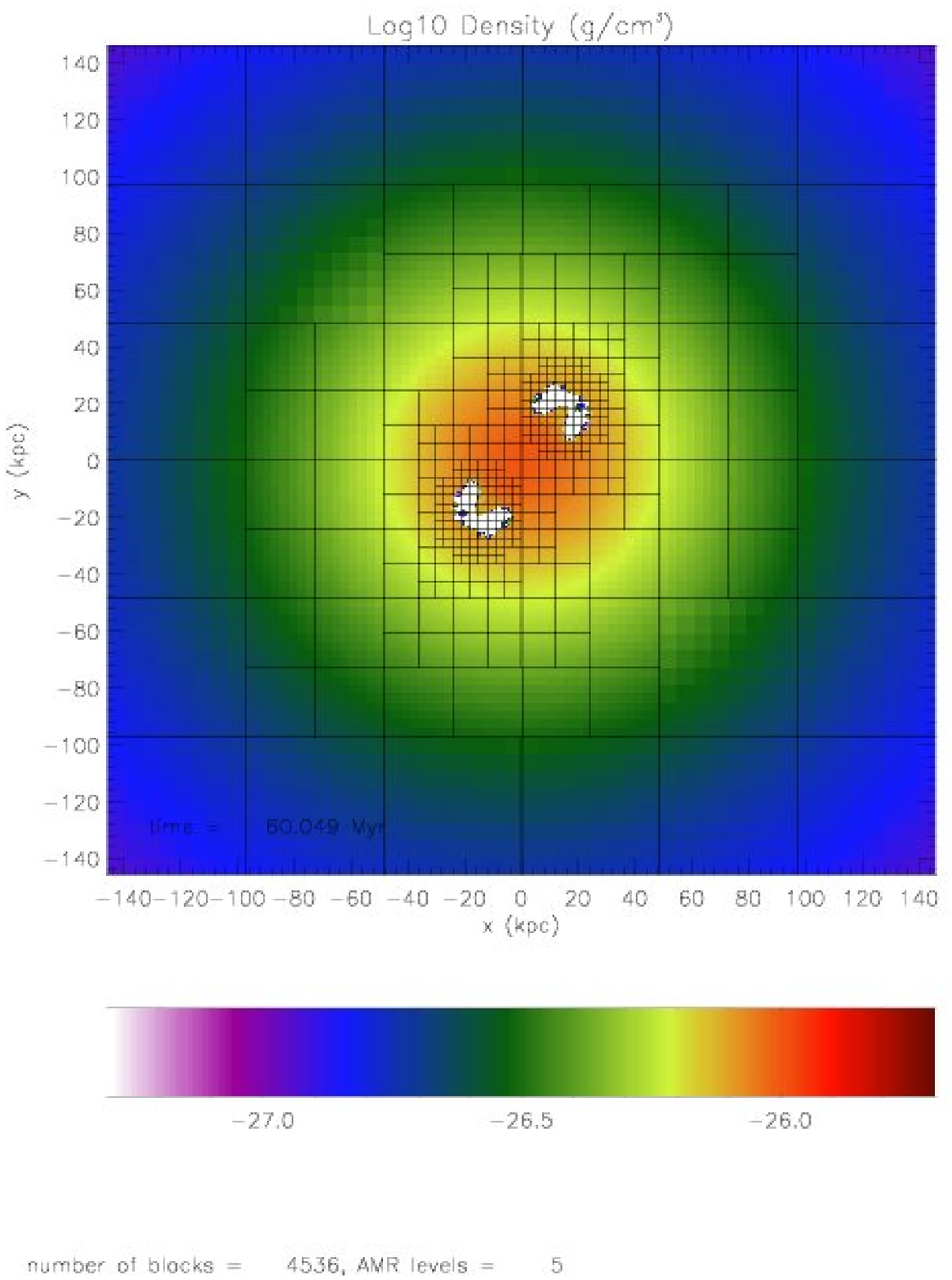}
\includegraphics[trim=0 80 0 0,clip,width=0.45\textwidth]{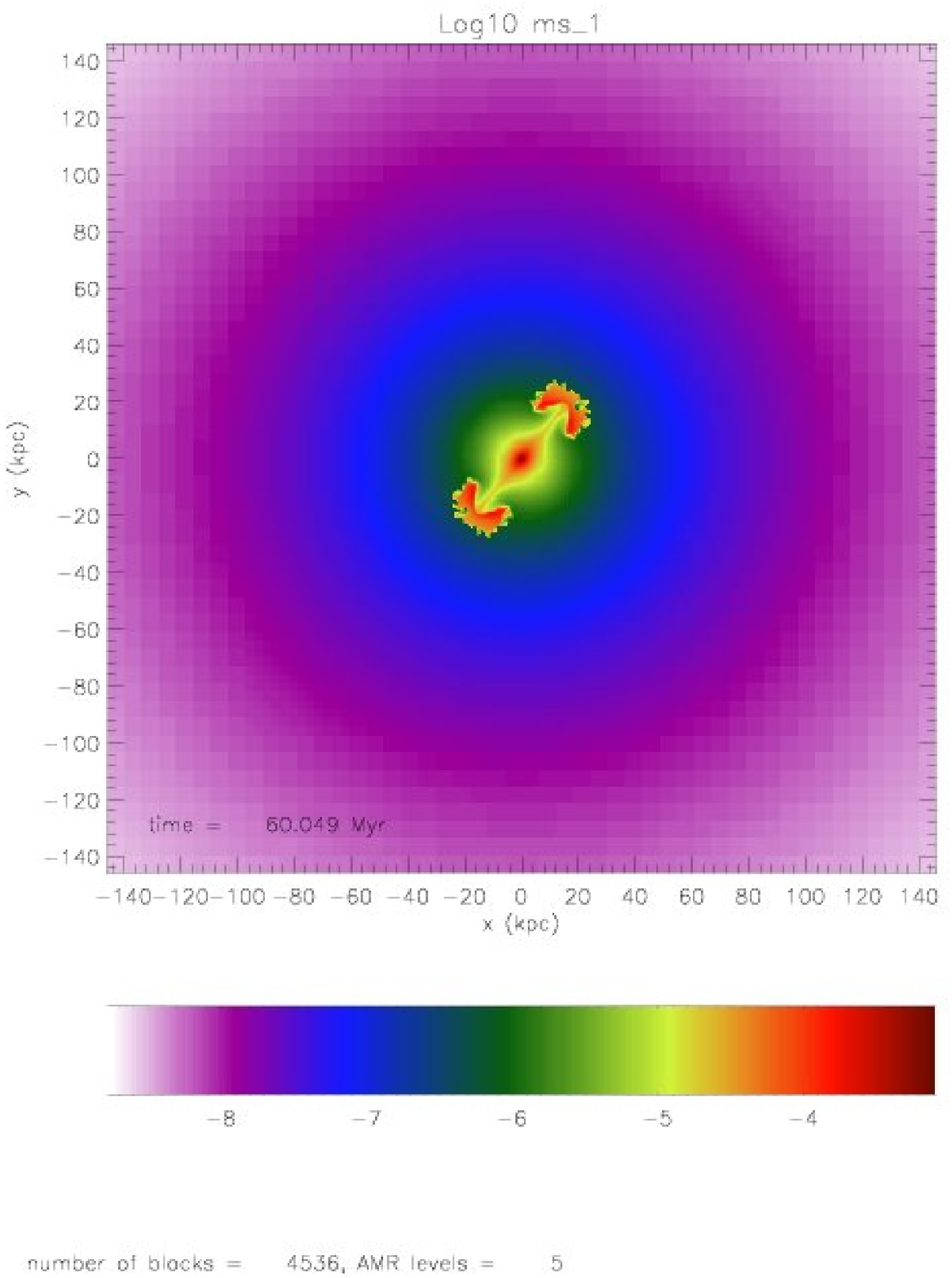}\newline
\includegraphics[trim=0 80 0 0,clip,width=0.45\textwidth]{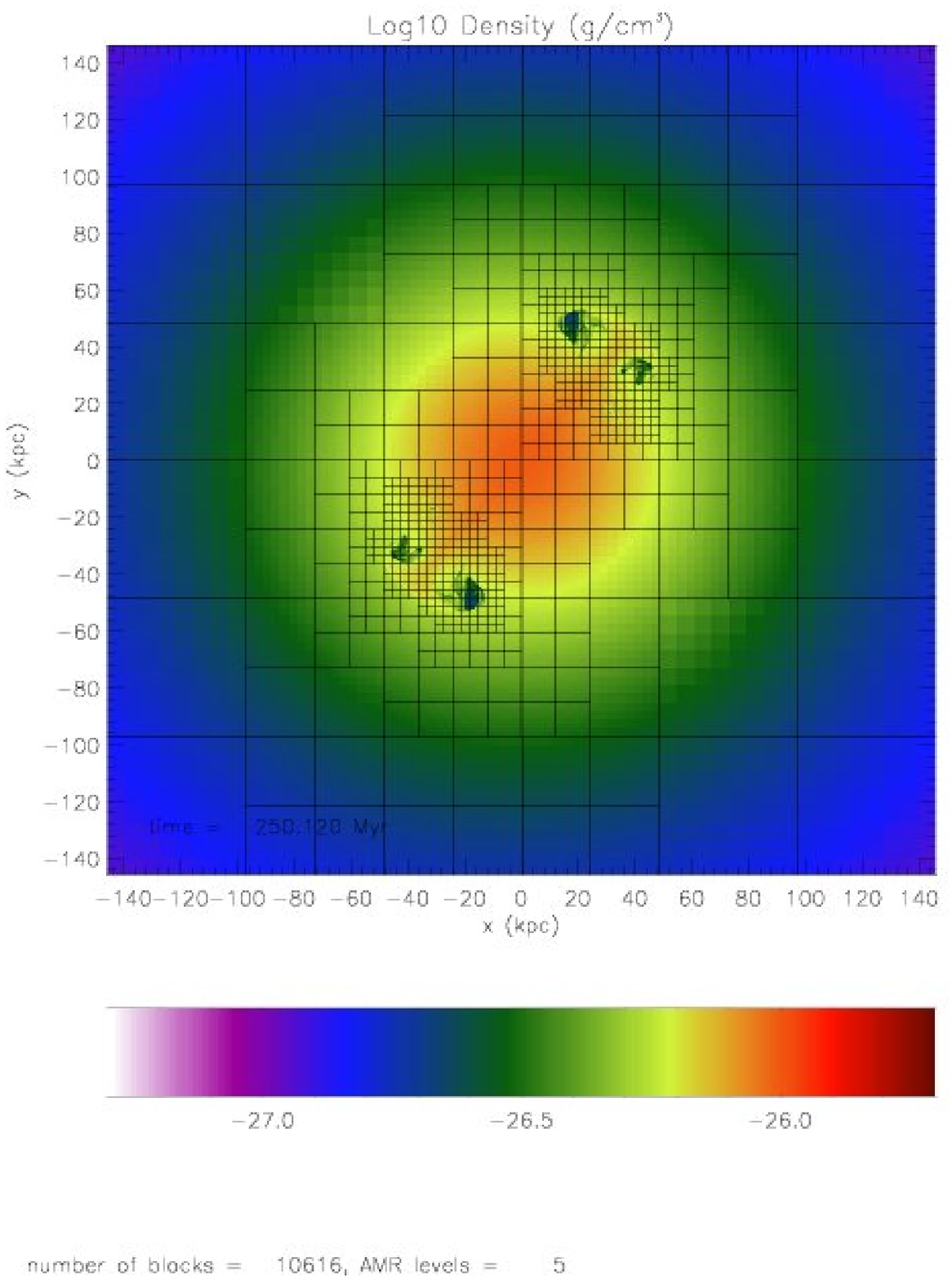}
\includegraphics[trim=0 80 0 0,clip,width=0.45\textwidth]{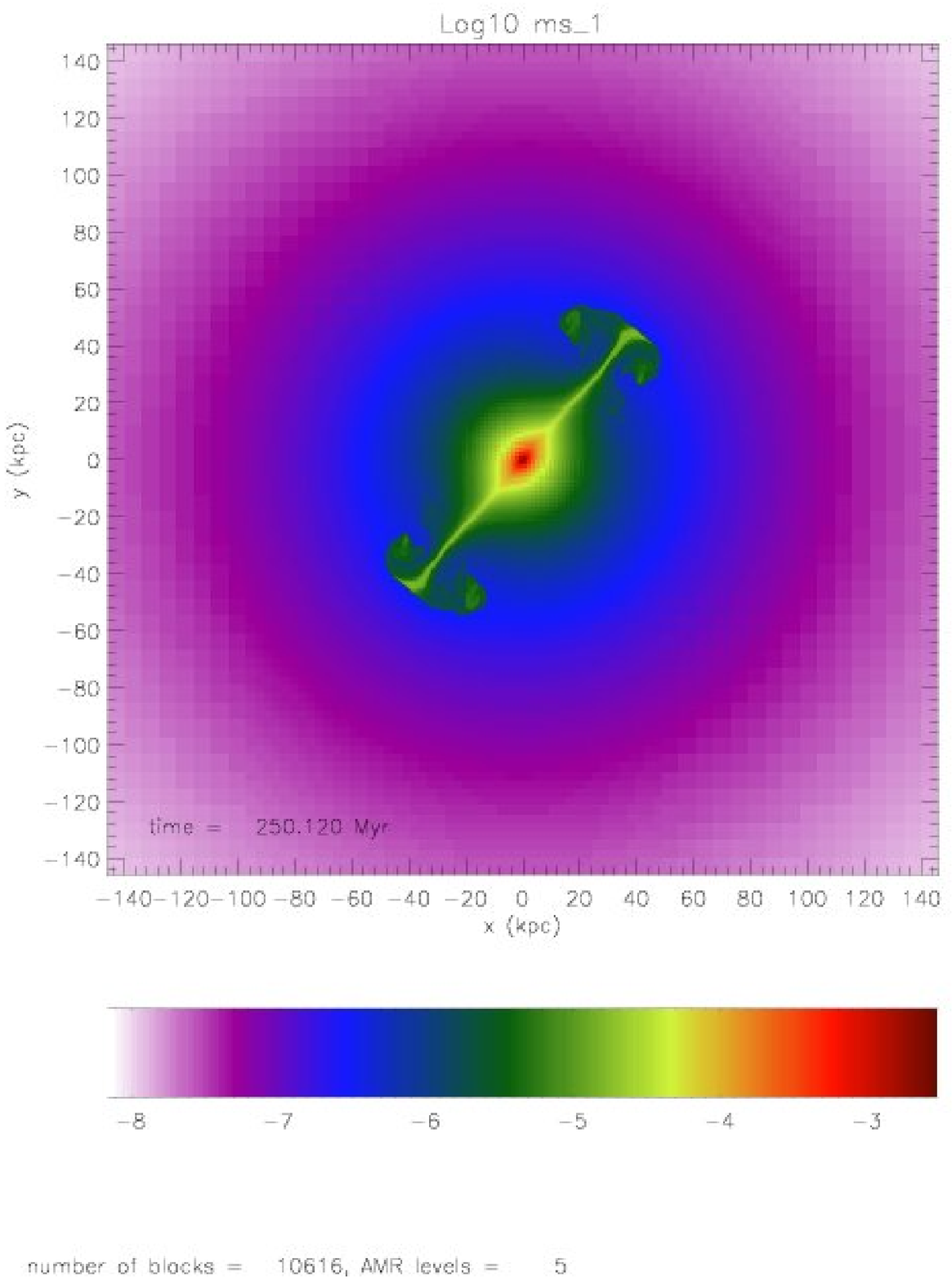}
\caption{Simulation of a single bubble pair: Slice through the centre
of the computational box showing the local gas density (left column)
  and the metal fraction (right column) at 60 Myr (top) and 250 Myr (bottom). The grid lines
  indicate the boundaries of the blocks. Each block contains $8^3$
  computational cells.}
\label{fig:slice_single_HR}
\end{figure*}

\begin{figure*}
\includegraphics[angle=-90]{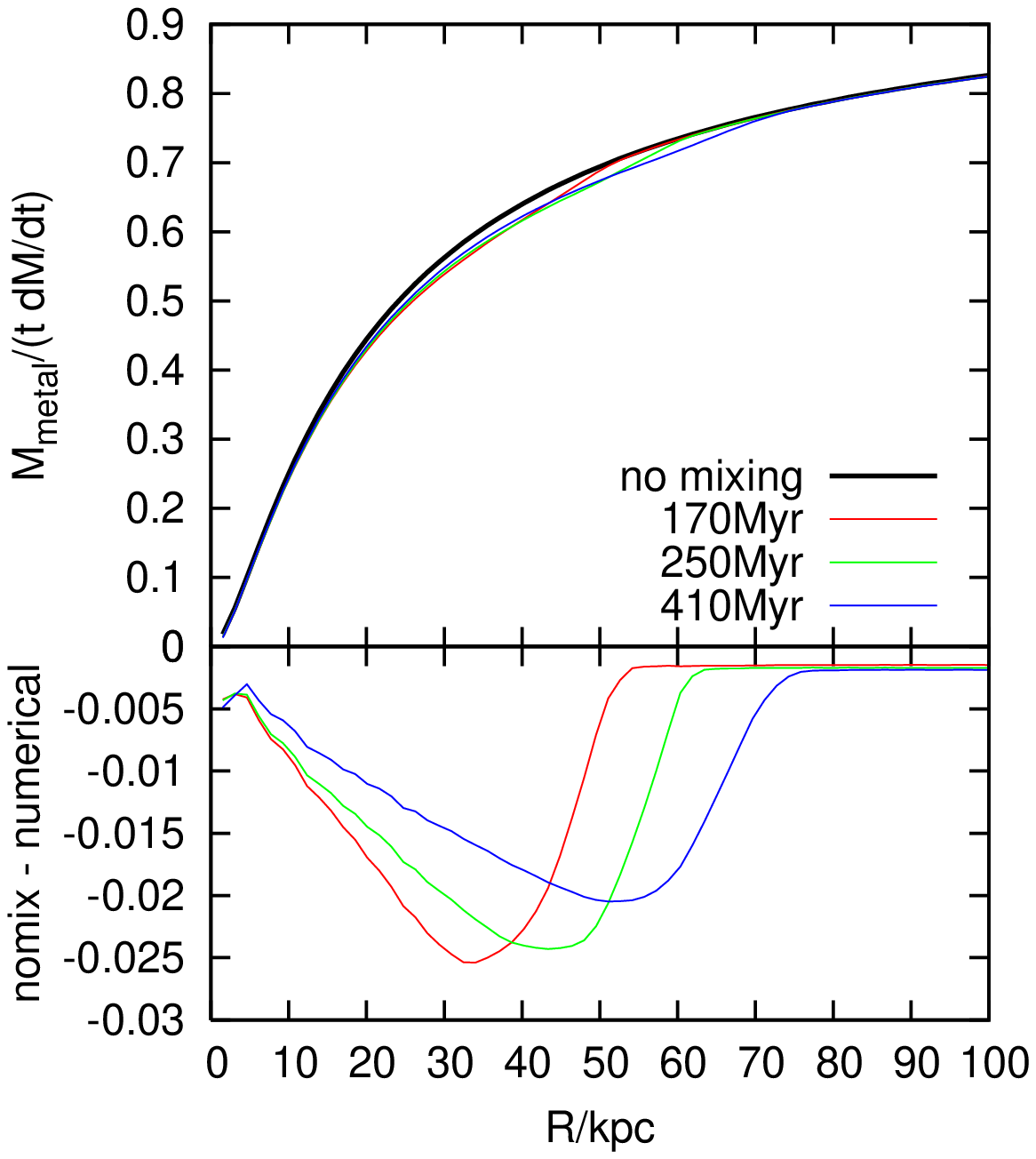}
\caption{Normalised cumulative metal mass for the run with a single
  pair of bubbles. These masses are averages in shells centred on the
  cluster. The thick solid line shows the cumulative metal mass
  without mixing. The bottom panel shows the difference between cumulative mass with and without mixing.}
\label{fig:prof_single_HR}
\end{figure*}

In the next step, we produced a single pair of bubbles by evacuating
two spherical regions of radius 10 kpc off-centre and on opposite
sides of the cluster centre. The distance of the bubble centre from
the cluster centre is 16 kpc. We studied the effect of this single
pair of bubbles on the metal profile in the cluster. In particular, we
wanted to compute how much mixing a single pair of bubbles can
cause and what the long term fate of these bubbles
is. Figures~\ref{fig:slice_single_HR} to \ref{fig:prof_single_HR} show
the results from this run.

As found in previous studies, the bubbles start to rise buoyantly and
fragment. While rising, they uplift the metals from the centre and
distribute them in their wake. Quite quickly the bubble fragments into
a torus. Associated with this torus is a vortical flow that causes gas
to ascend through the axis of the torus. Along the axis of the torus,
metals seem to be transported rather fast. However, this metal
spreading seems to be restricted to a narrow region around this
direction. Spreading along other directions seems negligible. The
effect is that the metals are spread out significantly along the axis
of the bubbles leading to a very elongated metal
distribution. Globally, the effect of a single pair of bubbles is not
very significant (see cumulative metal mass profile in
Fig.~\ref{fig:prof_single_HR}).\\

The bubbles do not continue to rise indefinitely and even after 1 Gyr
they do not rise beyond 100 kpc from the centre. When they fragment,
their rise velocity decreases. This can easily be seen by equating the
drag force to the buoyance force, i.e.

\begin{equation}
\frac{4}{3}\pi r^3\rho g = \frac{\pi}{2}Cr^2\rho v^2 \ ,
\end{equation}
where $v$ is the terminal velocity, $\rho$ the density of the ambient
medium, $g$ the gravitational acceleration, $r$ the radius of the
bubble and $C$ a constant of order unity. This yields the terminal velocity 

\begin{equation}
v^2= \frac{8rg}{3C} \ .
\label{eq:rise_vel}
\end{equation}

Thus, the rise velocity $\propto
\sqrt{r\Bubble}$ (see e.g. \cite{churazov:01}). The fragments form
a mushroom-shaped plume, not unlike the radio structures observed in
M87, and this structure covers a larger azimutal angle. There could be
a certain radius where most metals are deposited.

Two effects can lead to a more isotropic mixing of metals: (i) ambient
motions in the cluster that are triggered by galaxy motions and merger
activity and (ii) a series of bubbles that are launched at changing
locations in the cluster, for example by a precessing jet or by
inhomogeneities of the accretion flow near the jet. In the next
section, we investigate the second scenario.

\subsection{Series of bubbles}

In the next set of simulations, we look at the effect of a succession
of bubbles that are generated at changing positions inside the
cluster.

\subsubsection{Distance of bubbles from cluster centre}

\begin{figure*}
\includegraphics[angle=-90,width=0.4\textwidth]{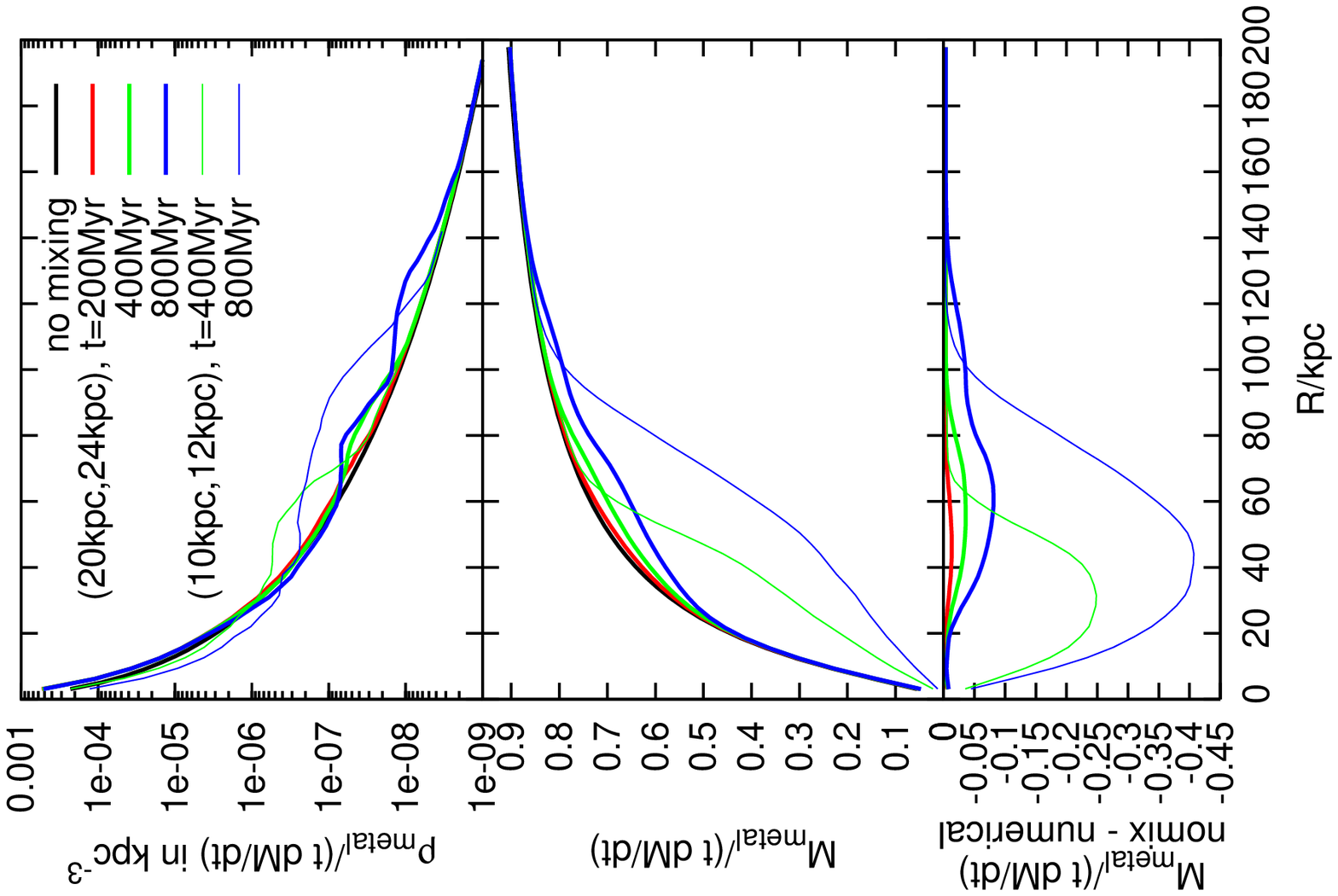}
\caption{Evolution of the normalised metal density profile (top panel) and the
  normalised cumulative metal mass profile (metal mass inside $r$, middle
  panel). The bottom panel shows the difference between mixing and no
  mixing. This figure shows the impact of the launch position of the bubbles. We compare runs EVAC and
  DISTCTR.}
\label{fig:prof_Mmets_distctr}
\end{figure*}

One would expect that the starting position of the bubbles with
respect to the central galaxy affects the amount of metals that can be
uplifted from the galaxy. This is indeed the case as we see by
comparing runs EVAC and DISTCTR. If bubbles start too far out, they
cannot uplift a substantial amount of metals because most metals are
injected mainly in the cluster centre (steep metal injection
profile) as can be seen from Fig.~\ref{fig:prof_Mmets_distctr}.

In order to compare metal fraction, metal density and cumulative metal
mass at different times, these quantities are normalised to
$\dot{M}\Metal{}_0 \cdot t$. From this we can conclude that the initial
position of the bubbles with respect to the scale radius of the metal
injection rate has an impact on the mixing efficiency of the
bubbles. However, as the bubbles are produced by jets that are launched from
supermassive black holes at the centre of the galaxy, the bubbles will
never be inflated too far from the stars that produce the
metals. While in principle the mixing efficiency depends on the
initial position of the bubble, the nature of the bubbles will ensure
that this will not cause great differences in the metal profiles.
Our generic run has a distance of the bubble centre from the core of
the cluster of $\sim 16$ kpc. In reality, bubbles occur at a broad range
of distances from the centre. The list of known bubbles in
\citet{dunn:05} shows that the distances that we probe lie well within
the observed ones.

\subsubsection{Bubble size}

\begin{figure*}
\includegraphics[angle=-90,width=0.4\textwidth]{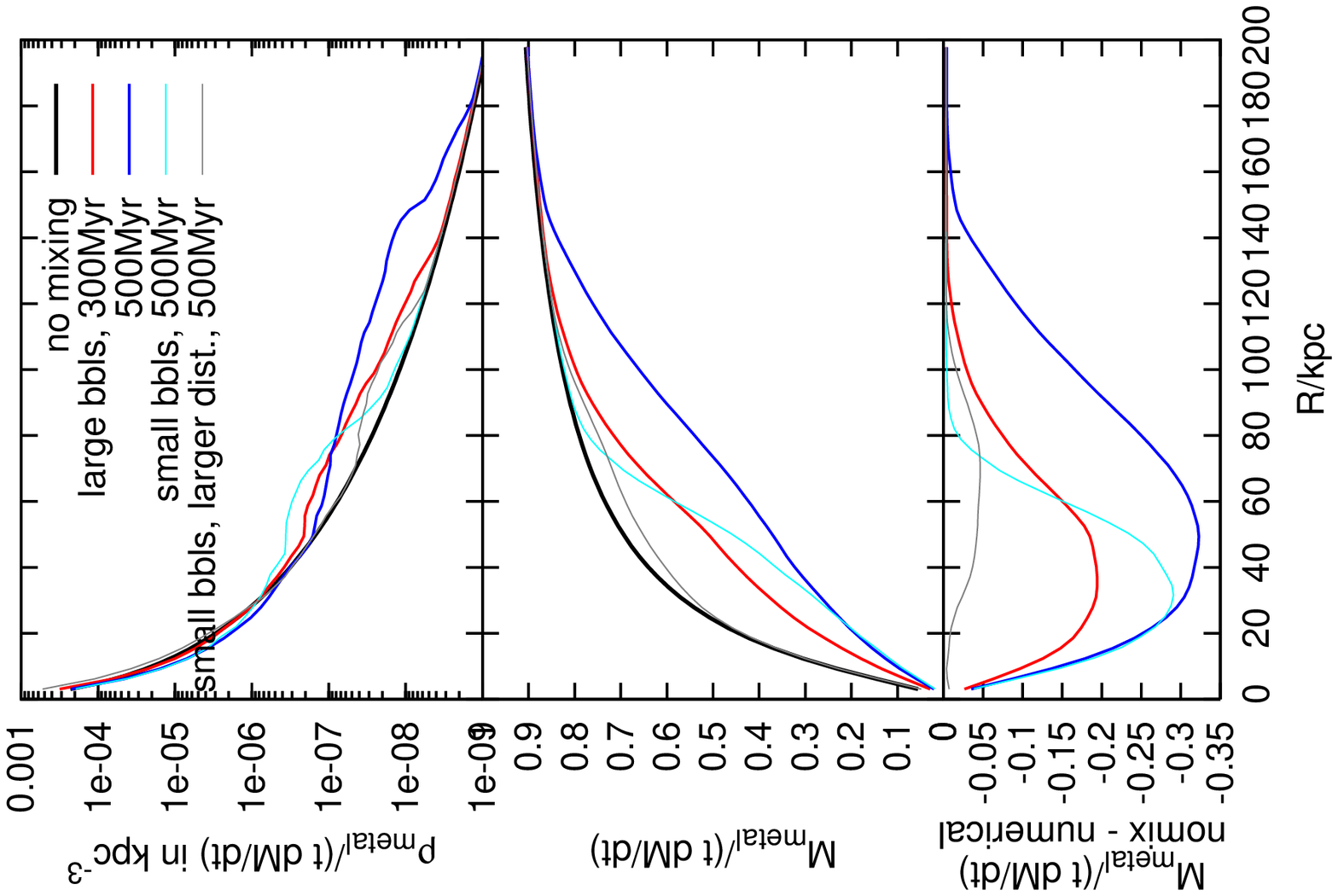}
\caption{Evolution of the normalised metal density profile (top panel) and the
  normalised cumulative metal mass profile (metal mass inside $r$, middle
  panel). The bottom panel shows the difference between mixing and no
  mixing. Here we illustrate
the impact of the initial bubble position and the initial bubble
  sizes. Shown are results from runs EVAC,
  DISTCTR and LARGEBBL.}

\label{fig:prof_Mmets_large}
\end{figure*}

Obviously, one important parameter is the size of the bubbles. In our
standard scenario we chose a bubble radius of 10 kpc which is
motivated by the sizes of some of the cavities observed in the X-ray
surface brightness maps. For example, the cavities of the inner
bubbles in Perseus have a radius of 13 kpc. To test the dependence of
the transport efficiency on the bubble size, we compared our standard
run with a bubble of twice its radius (8 times its volume). See
Fig.~\ref{fig:prof_Mmets_large} for the cumulative metal mass
profiles.  The main inference is that larger bubbles can carry metals
out to larger distances. As can be seen in
Fig.~\ref{fig:prof_Mmets_large}, after about 500 Myrs the large bubble
($r=20$ kpc) carries the metals out to a distance of about 50 kpc,
while the 10 kpc bubble only lifts them to about 25 kpc.

\begin{figure*}
\includegraphics[width=0.45\textwidth]{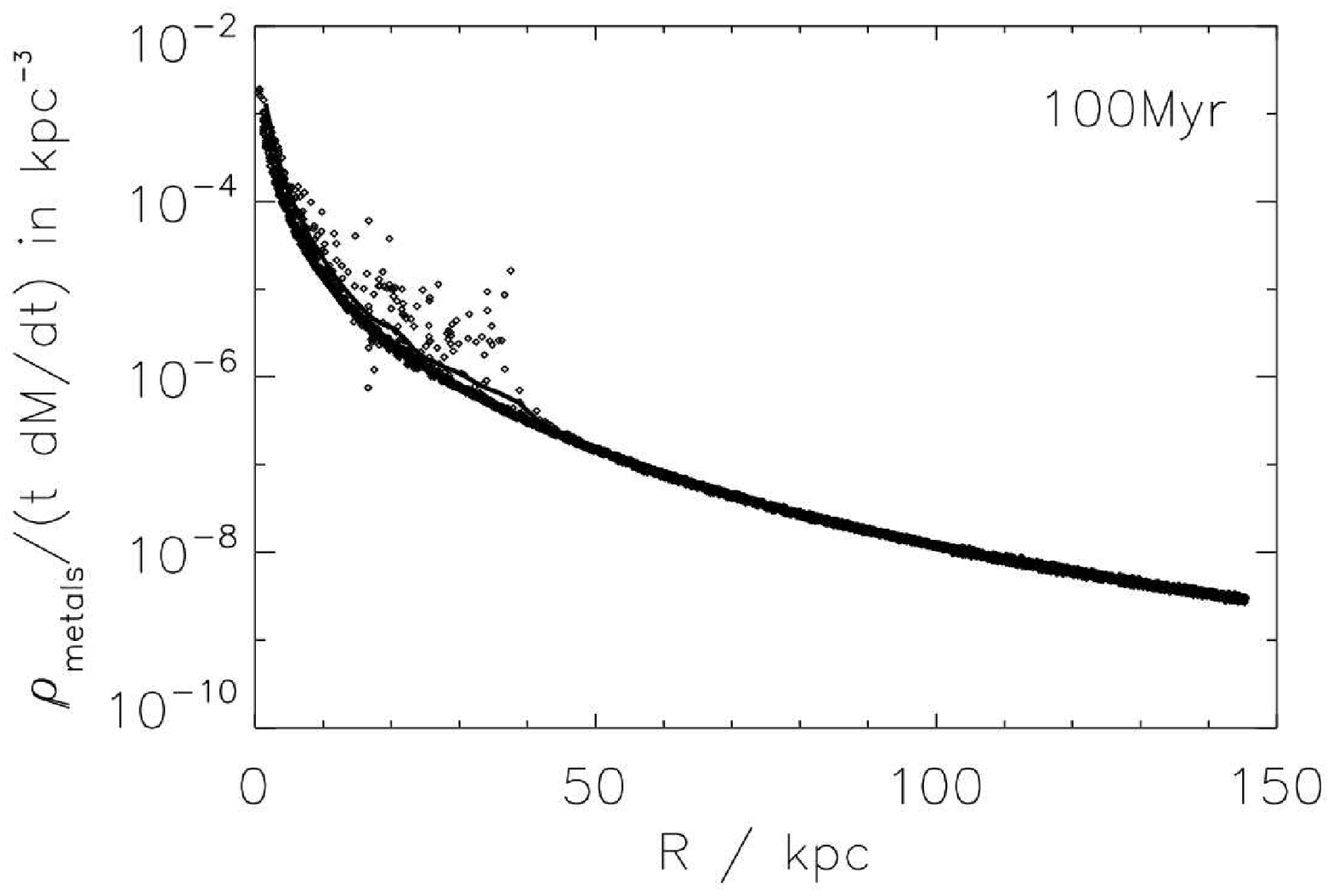}
\includegraphics[width=0.45\textwidth]{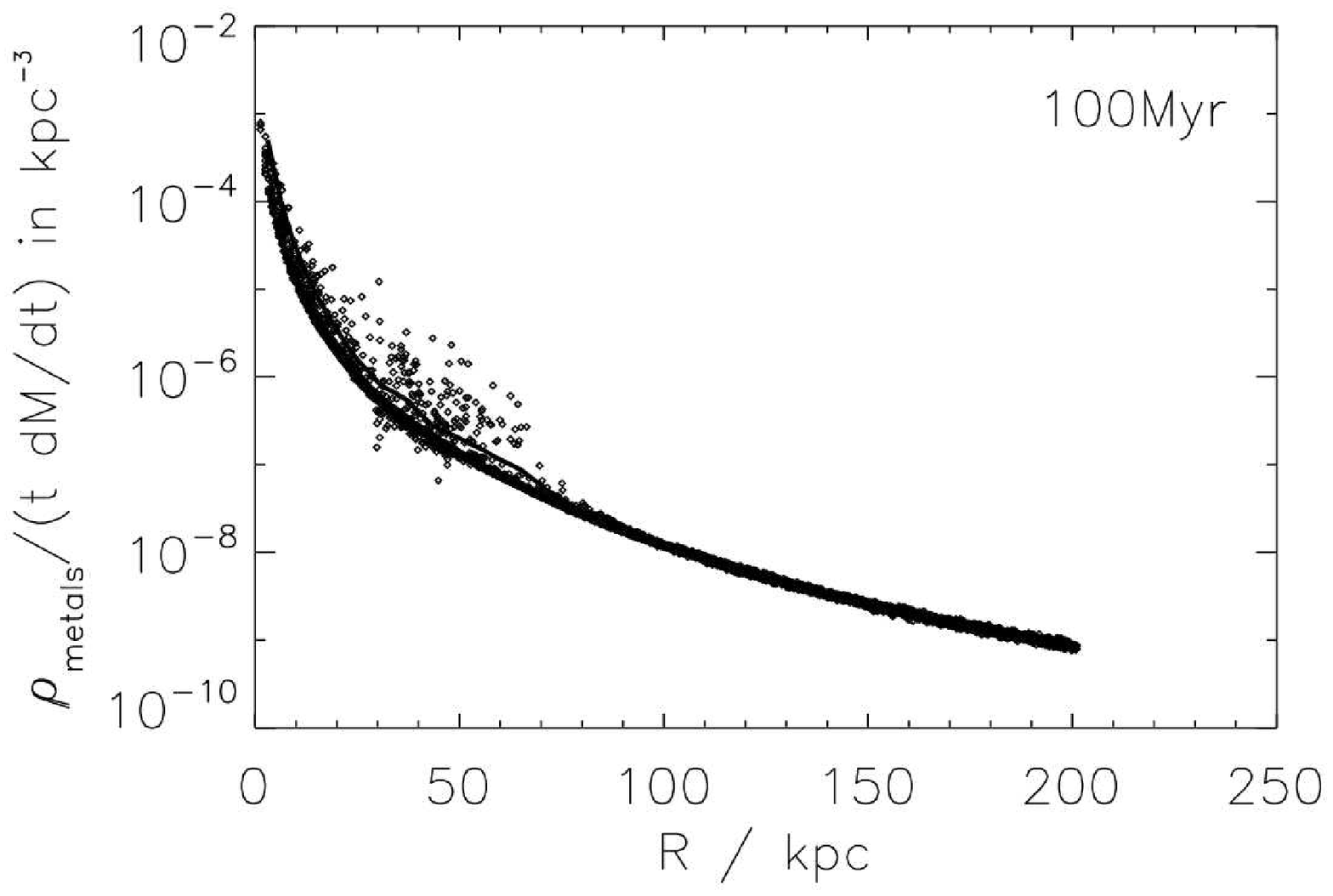}\vspace*{-0.5cm}\newline
\includegraphics[width=0.45\textwidth]{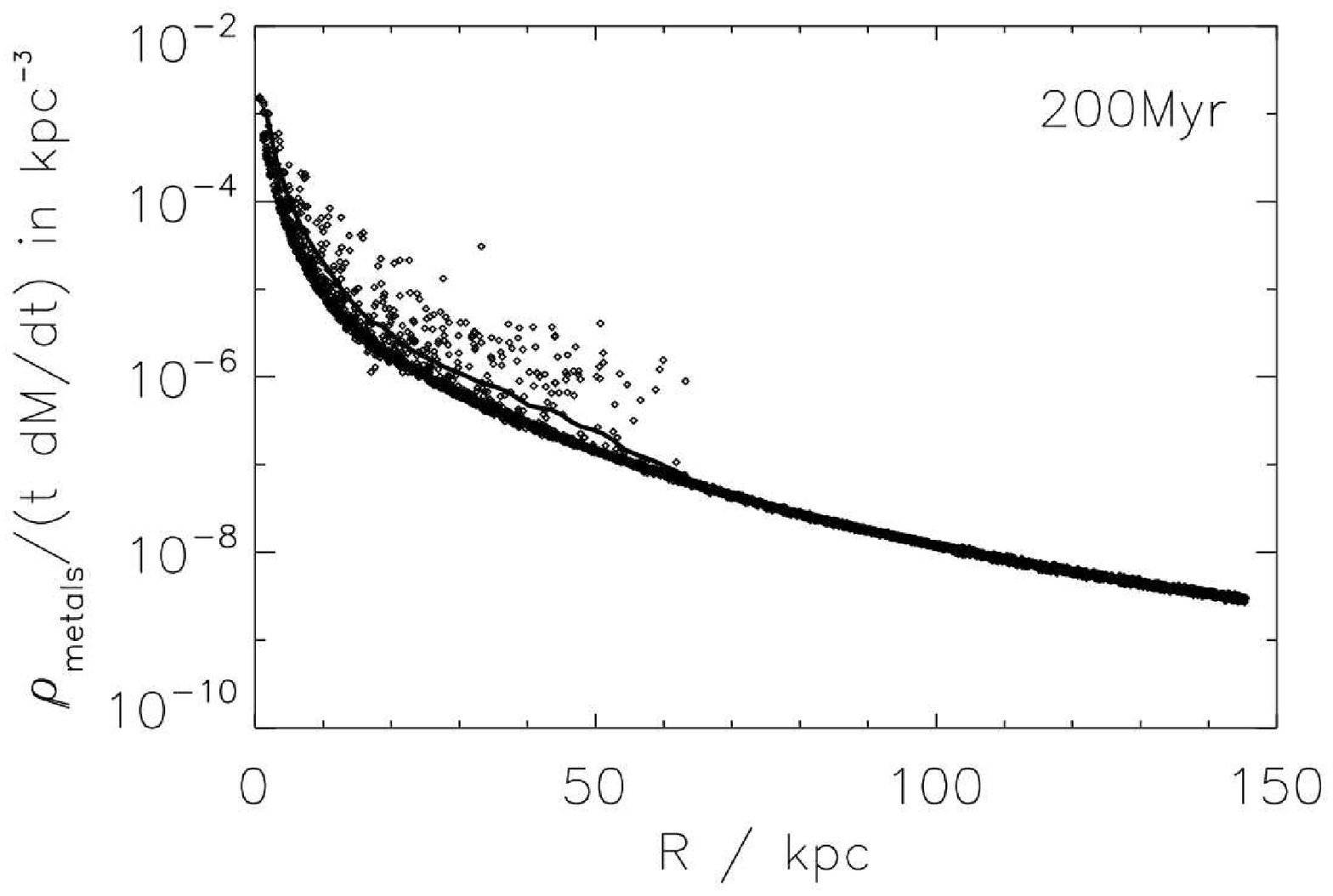}
\includegraphics[width=0.45\textwidth]{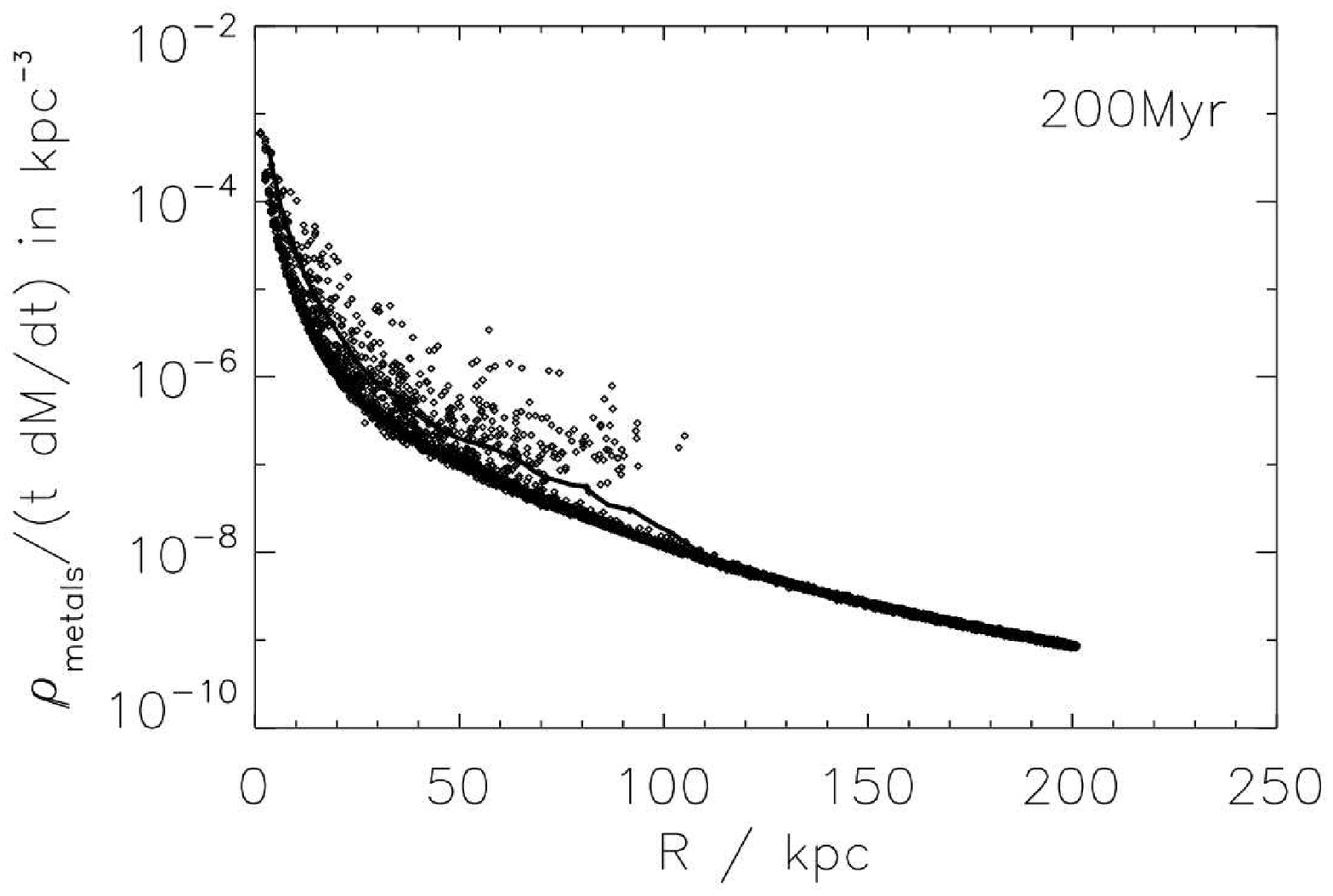}\vspace*{-0.5cm}\newline
\includegraphics[width=0.45\textwidth]{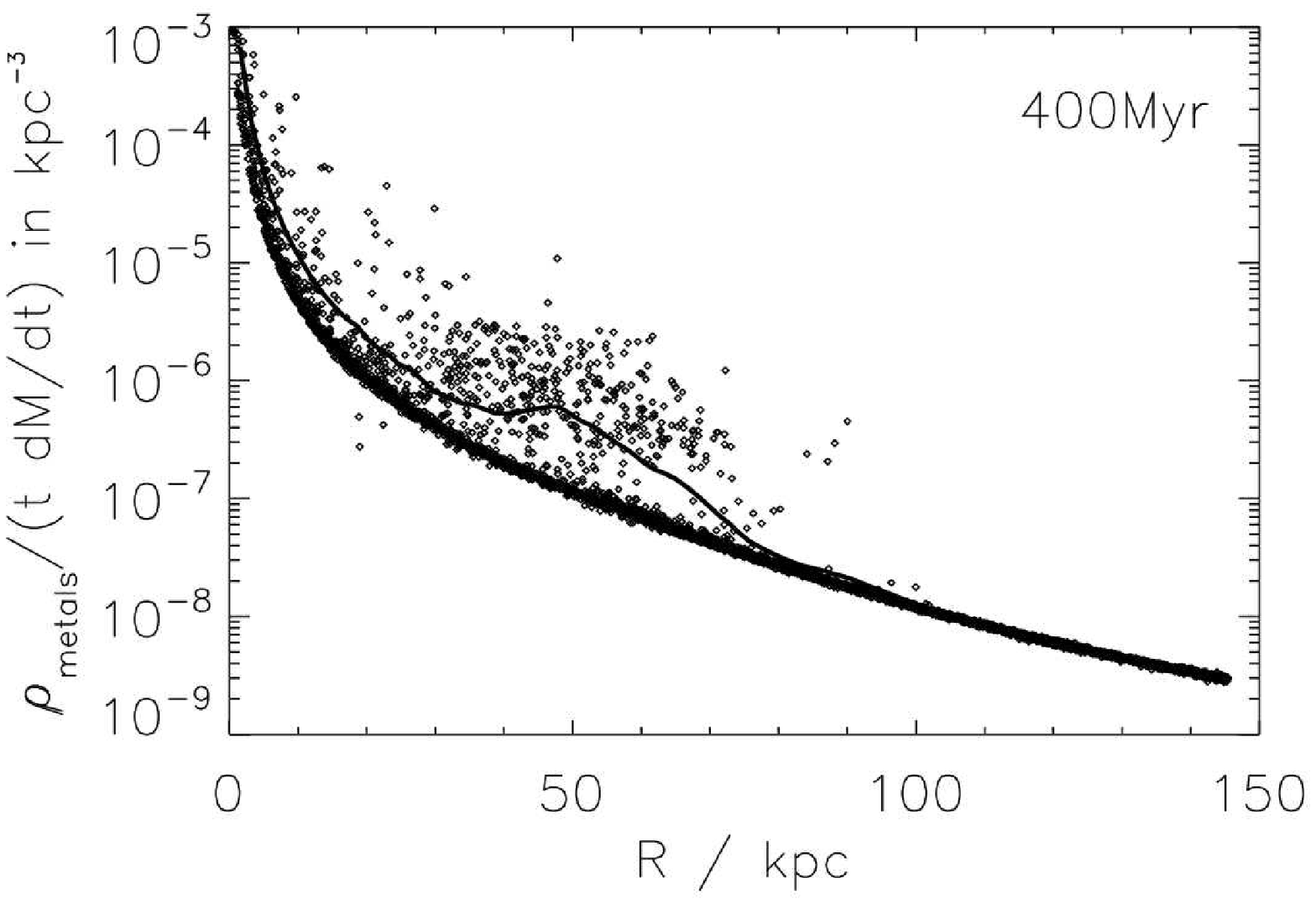}
\includegraphics[width=0.45\textwidth]{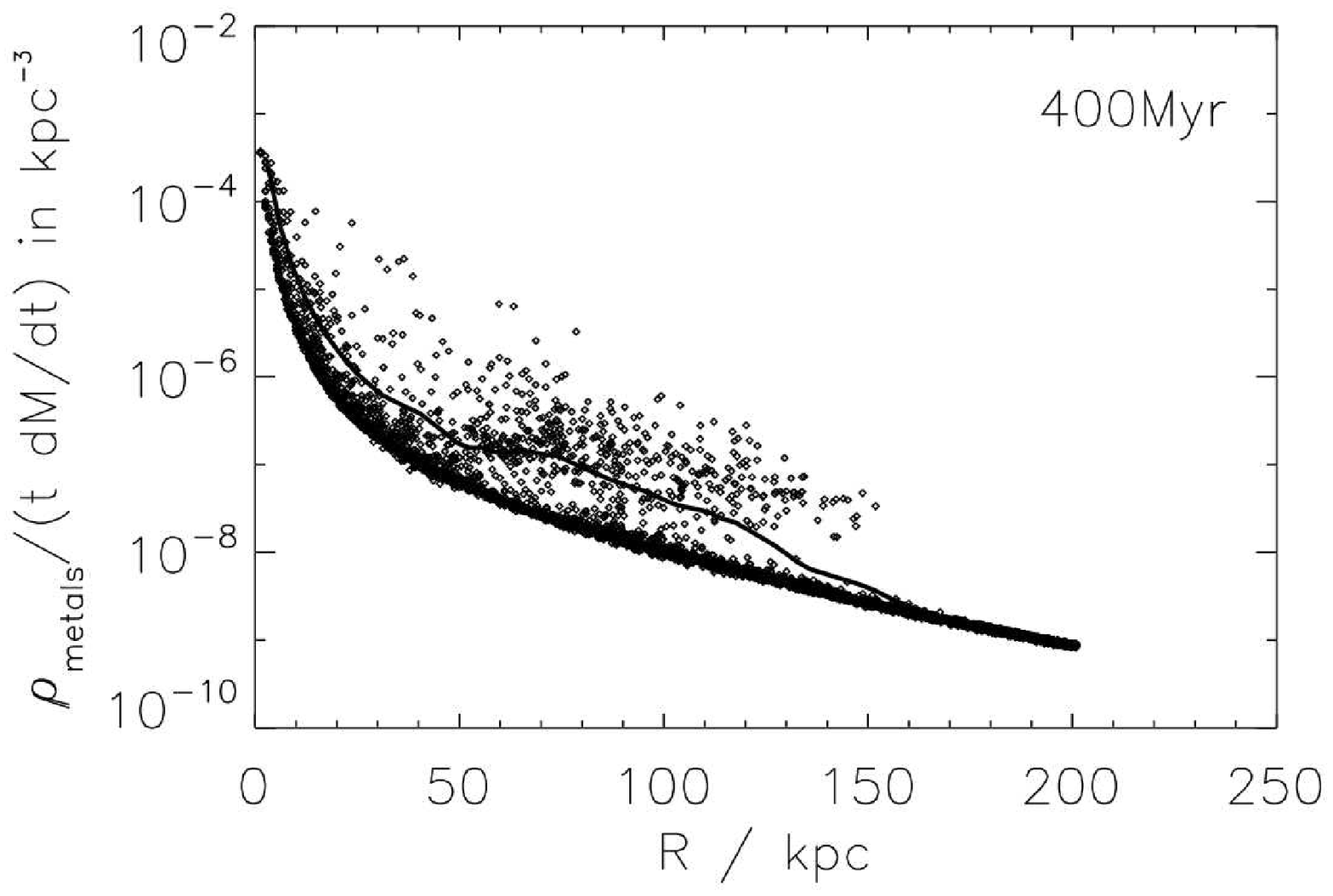}
\caption{Normalised radial metal density profile for three different
  times. A random subset of cells is
  chosen (same weight for each radius). Metal density is normalised to $t\dot
  M\Metal{}_0$. Left column is for EVAC, right column for
  LARGEBBL. The main effect of the bubbles can be seen in the
  points elevated above the original line. In these cells, the metal density is
  enhanced because the bubbles have carried the metals to larger
  radii. It is apparent that the larger bubbles are more efficient at
  transporting metals outwards.}

\label{fig:prof_rhomets_evac_infl}
\end{figure*}

The same trend is apparent from Fig.~\ref{fig:prof_rhomets_evac_infl}
  which shows the normalised radial metal density profile for three different
  times. The main effect of the bubbles can be seen in the
  points elevated above the original line. In these cells, the metal density is
  enhanced because the bubbles have carried the metals to larger
  radii. It is evident that the larger bubbles are more efficient at
  transporting metals outwards.

This behaviour is not surprising because larger bubbles rise faster as
we saw in Eq.~(\ref{eq:rise_vel}). As the bubbles rise faster, they
accelerate the metals in its wake to higher velocities and they travel
further out until instabilities have broken them up such that they do
not travel much further.

\subsubsection{Bubble frequency}

Clearly, the efficiency with which bubbles can transport metals
depends on the frequency with which bubbles are produced by the
AGN. In the previous simulations, we employed a recurrence time of 50
Myr. For comparison, we ran a simulation with a recurrence time of 200
Myrs, all other parameters being the same as in run EVAC.

\begin{figure*}
\includegraphics[angle=-90,width=0.4\textwidth]{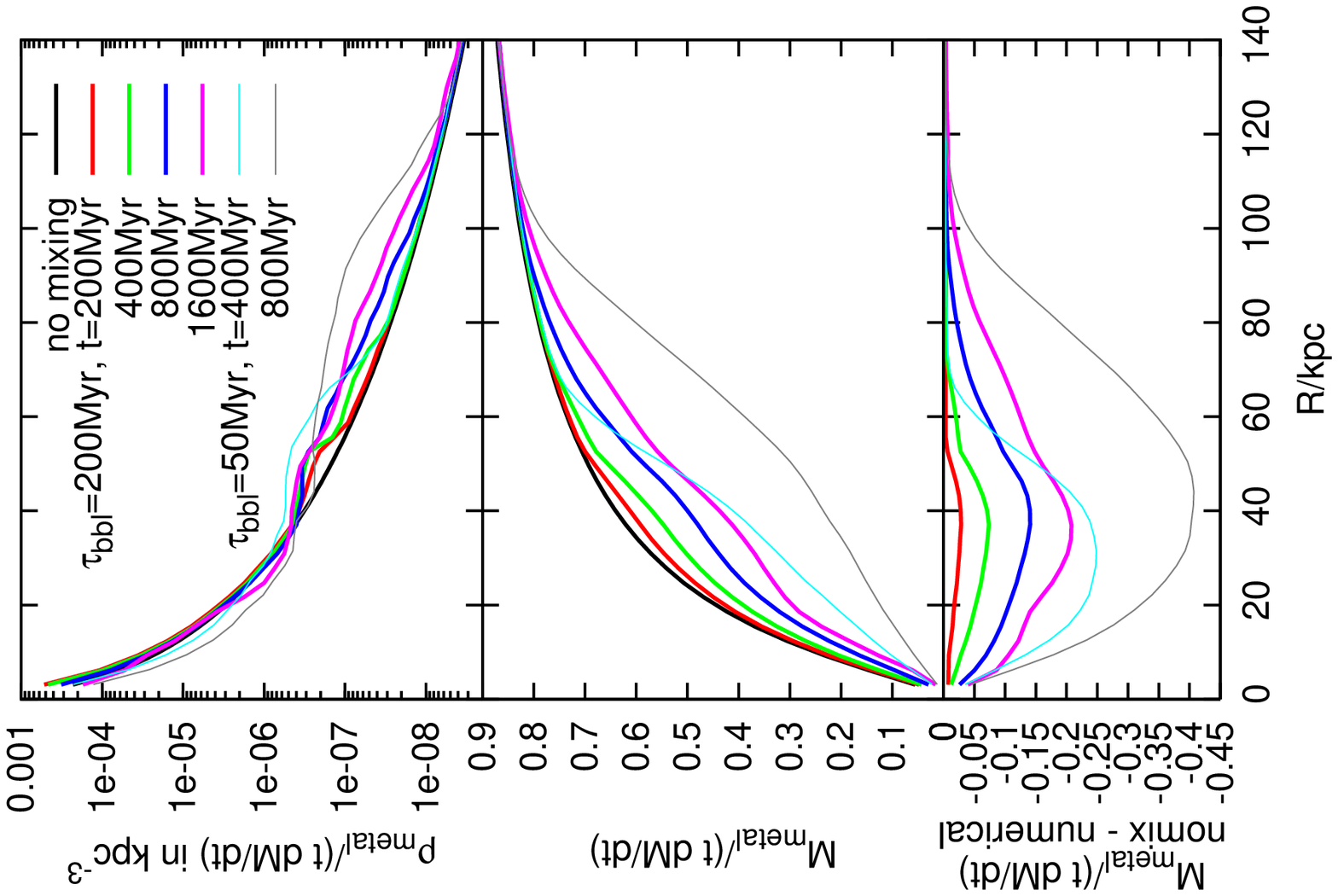}
\caption{Evolution of normalised metal density profile (top panel) and and
  normalised cumulative metal mass profile (metal mass inside $r$,
  middle panel). The bottom panel shows the difference between mixing
  and no mixing. Comparison of two runs with different bubble
  evacuation timescale ($\tau\Bubble$) (runs EVAC and TAUBBL200).
  Different line styles and colours correspond to different times and
  runs, see legend. The black line is the prediction of what should
  happen without any mixing.}
\label{fig:prof_Mmets_taubbl}
\end{figure*}

The difference in, both, radius up to which metals are carried and
degree of deformation of cumulative metal mass profile can be seen in
Fig.~\ref{fig:prof_Mmets_taubbl}. After 800 Myrs, comparing the blue
and grey lines reveals the difference between these two runs. The runs
with a longer recurrence time leads to appreciably less mixing. A
recurrence time that is four times longer, roughly leads to only about
1/3 as much mixing by the end of our simulation. Thus, the degree of mixing
scales roughly with the frequency of bubble generation as one would
expect.

\subsubsection{Comparison between evacuation and inflation method}

\begin{figure*}
\includegraphics[angle=-90,width=0.45\textwidth]{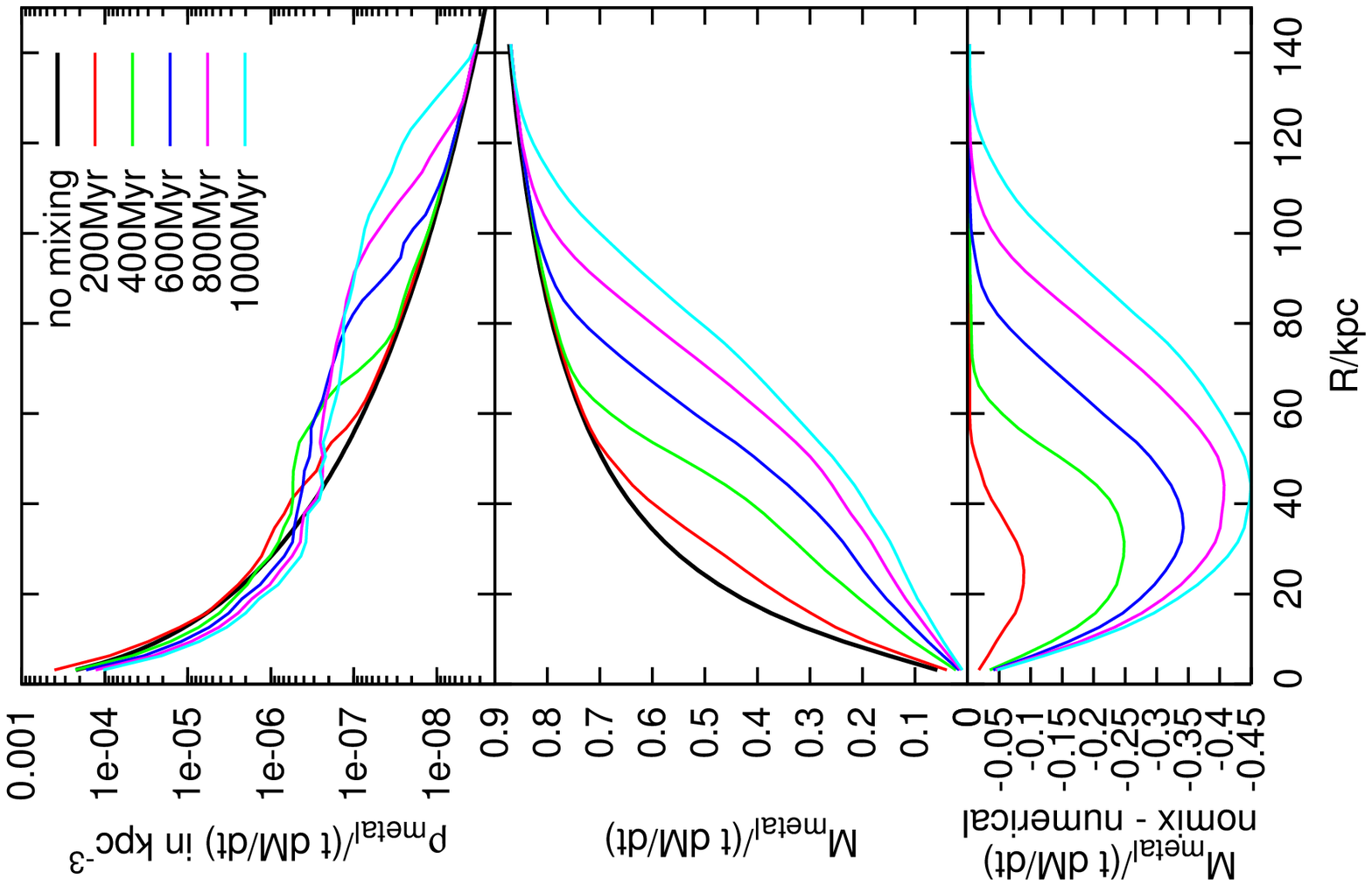}
\includegraphics[angle=-90,width=0.45\textwidth]{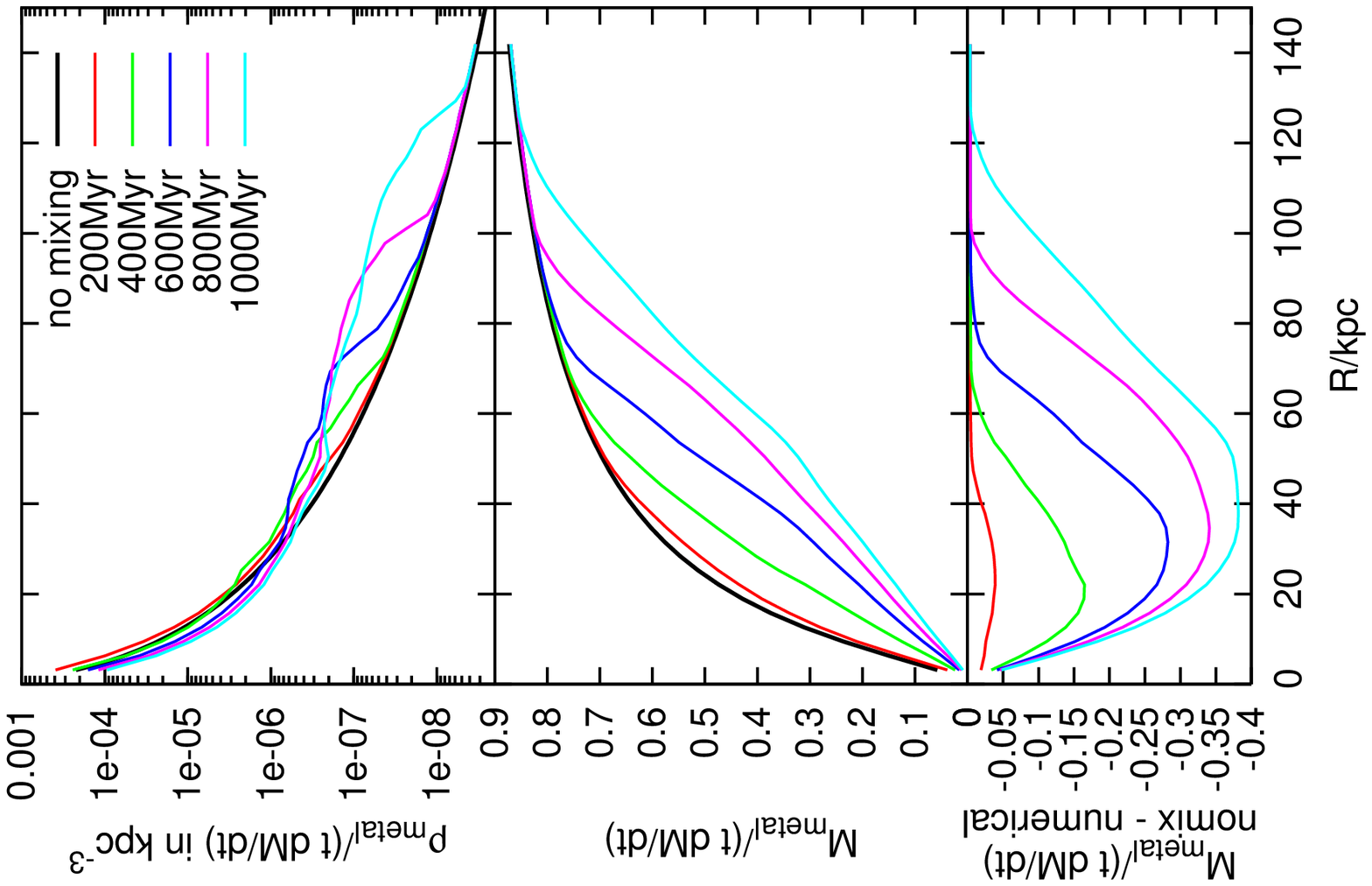}
\caption{Evolution of the metal density profile (top panels) and the cumulative
  metal mass profile (metal mass inside $r$, middle panels).  Both profiles
  are normalised to $t\dot M\Metal{}_0$. Different timesteps are colour-coded,
  see legend. Left is for bubble evacuation method, right for bubble inflation
  method. The black line is the prediction of what should happen without any
  mixing. The bottom panels show the difference between mixing and no mixing.}
\label{fig:prof_Mmets_evac_infl}
\end{figure*}

\begin{figure*}
\vspace*{-0.5cm}
\includegraphics[width=0.32\textwidth]{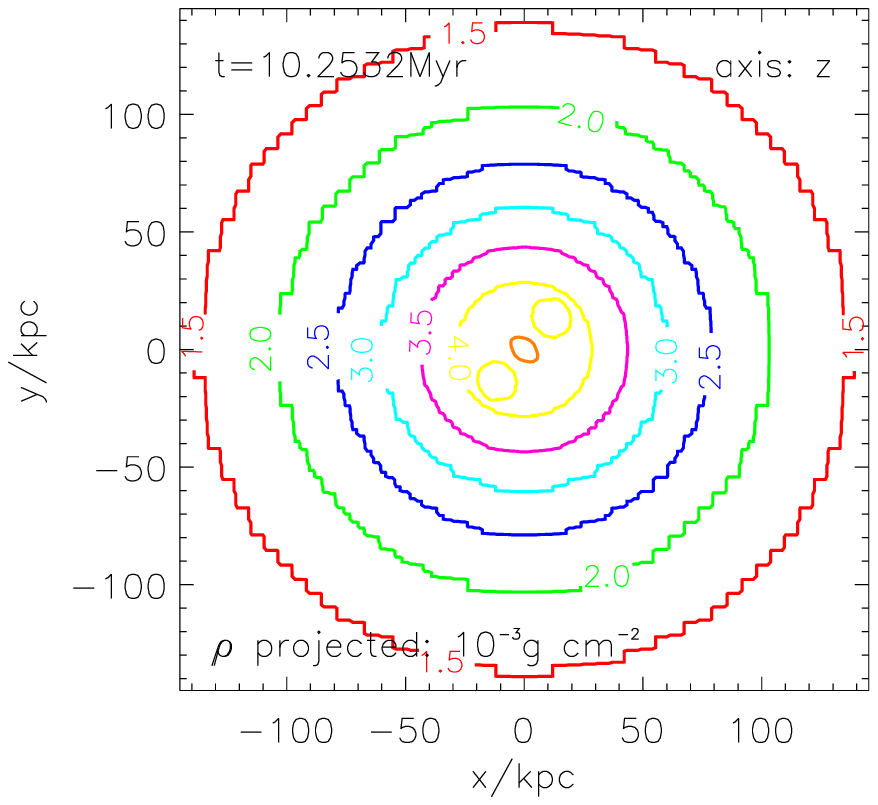}
\includegraphics[width=0.32\textwidth]{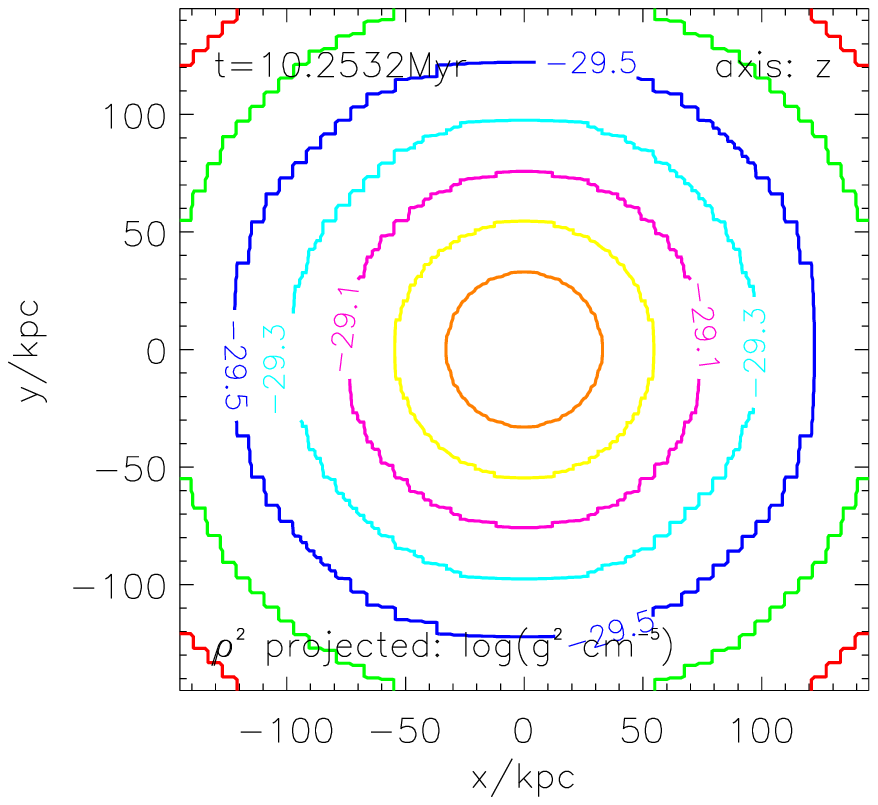}
\includegraphics[width=0.32\textwidth]{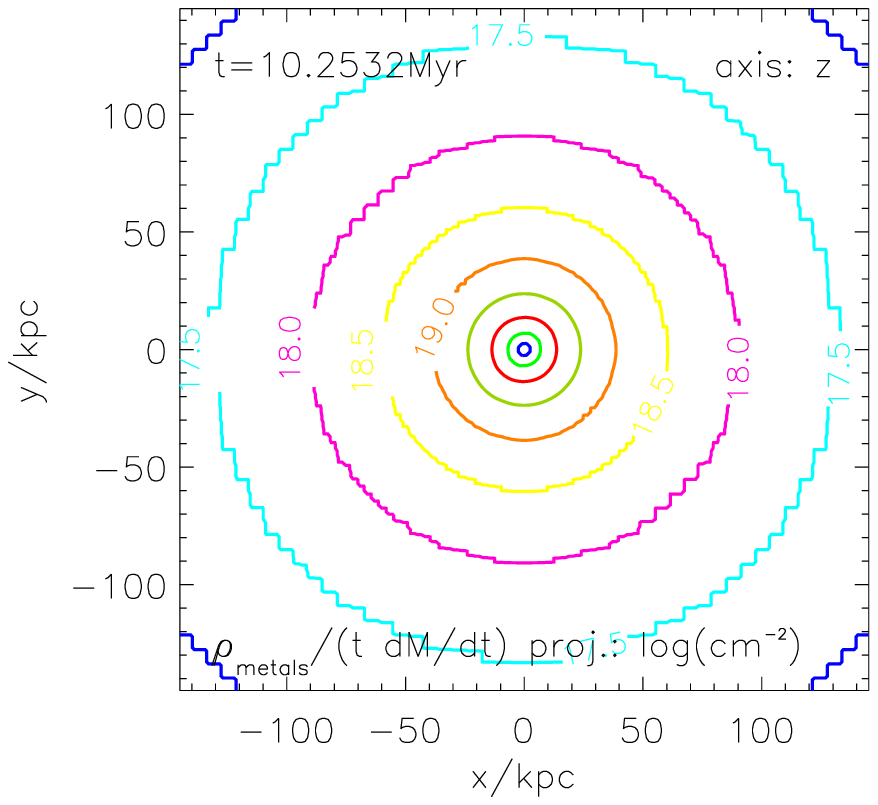}\vspace*{-1cm}\newline
\includegraphics[width=0.32\textwidth]{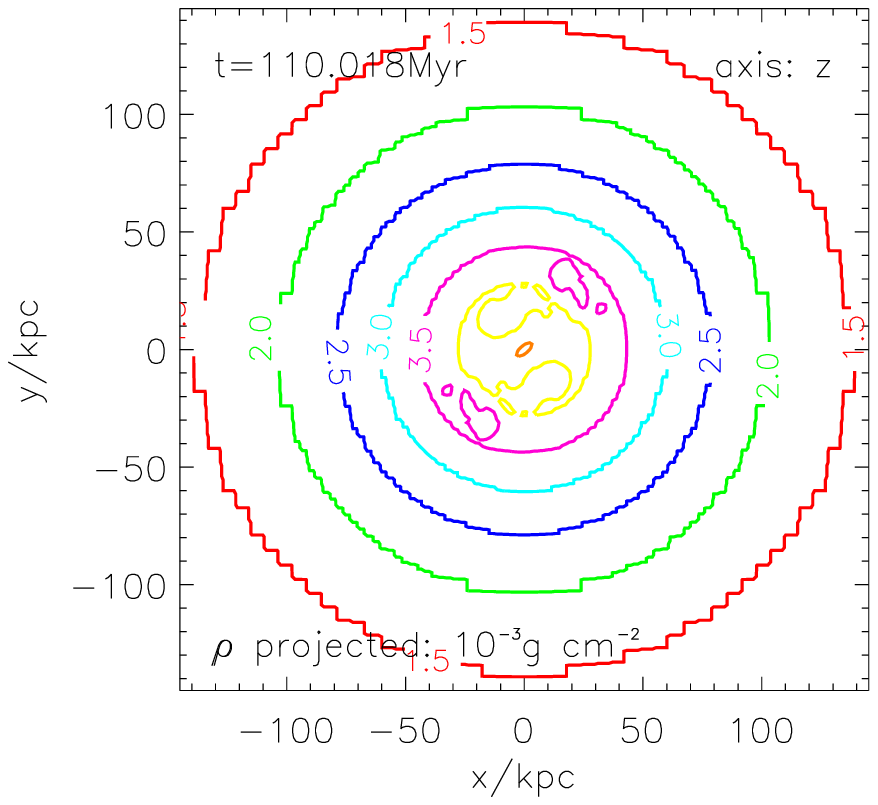}
\includegraphics[width=0.32\textwidth]{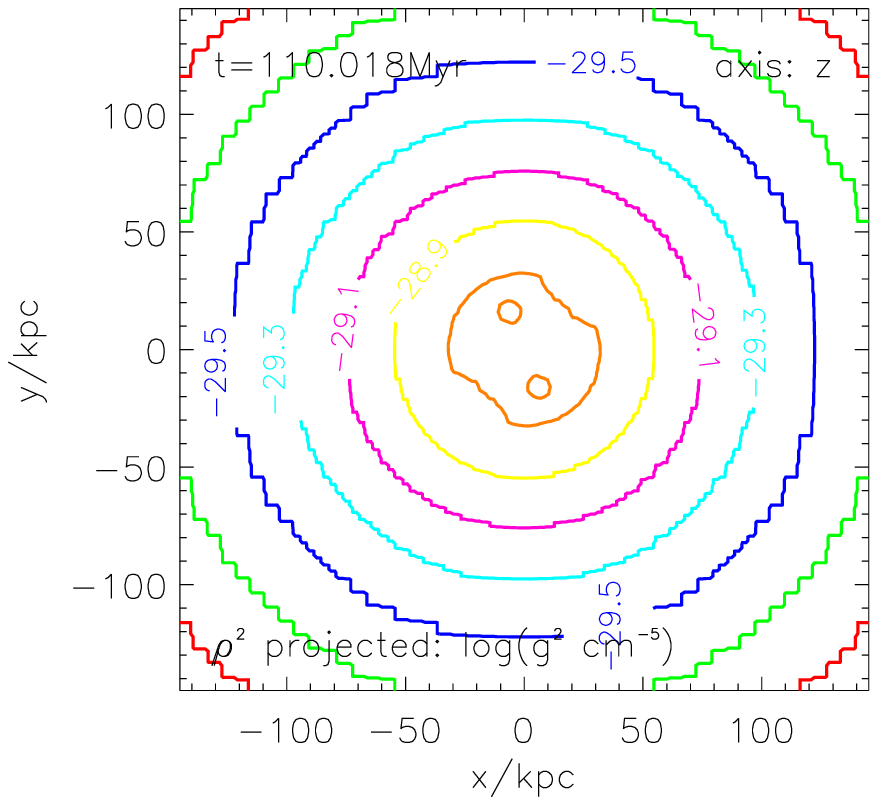}
\includegraphics[width=0.32\textwidth]{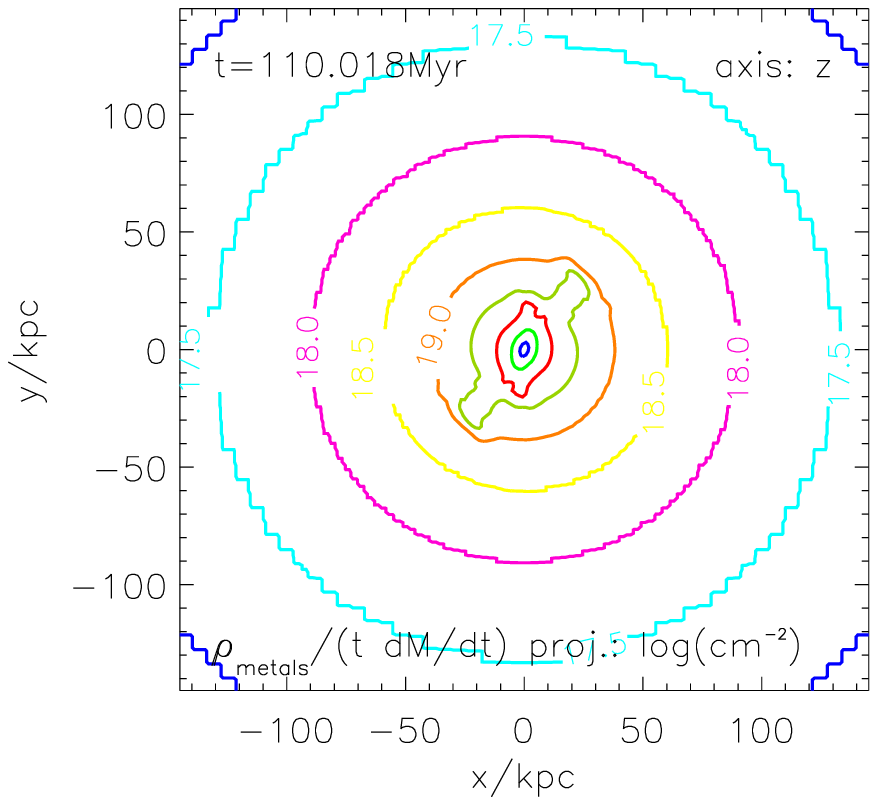}\vspace*{-1cm}\newline
\includegraphics[width=0.32\textwidth]{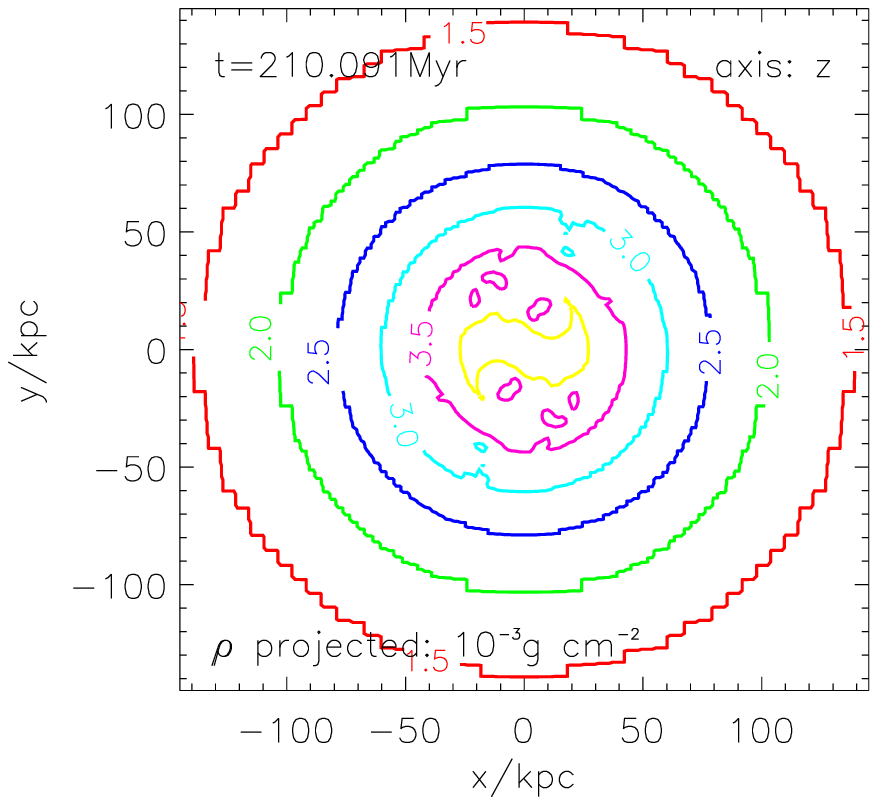}
\includegraphics[width=0.32\textwidth]{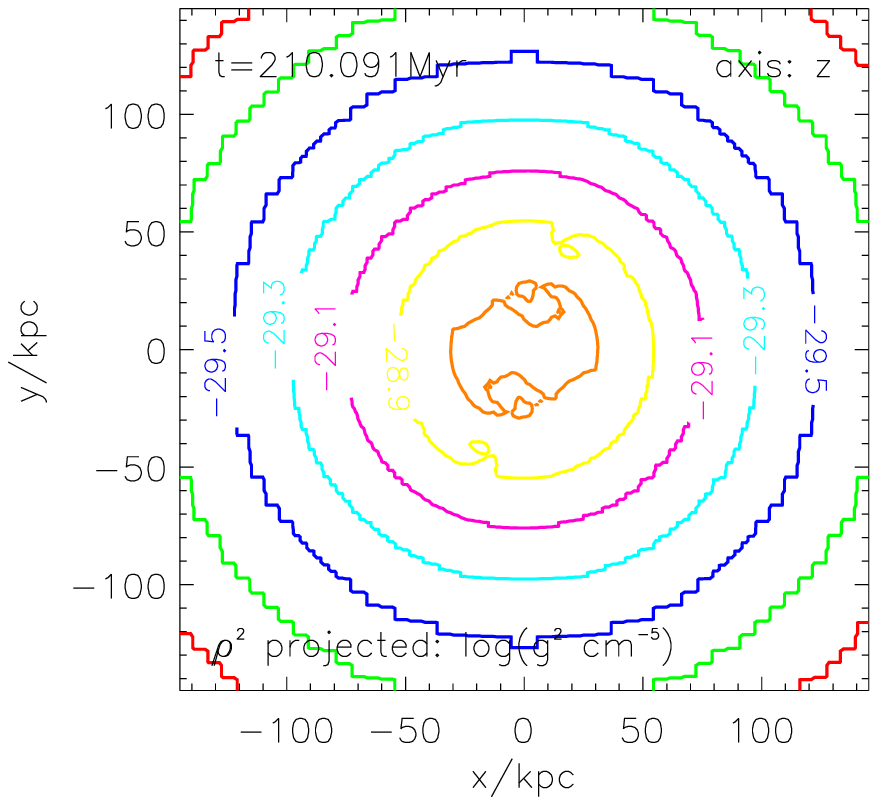}
\includegraphics[width=0.32\textwidth]{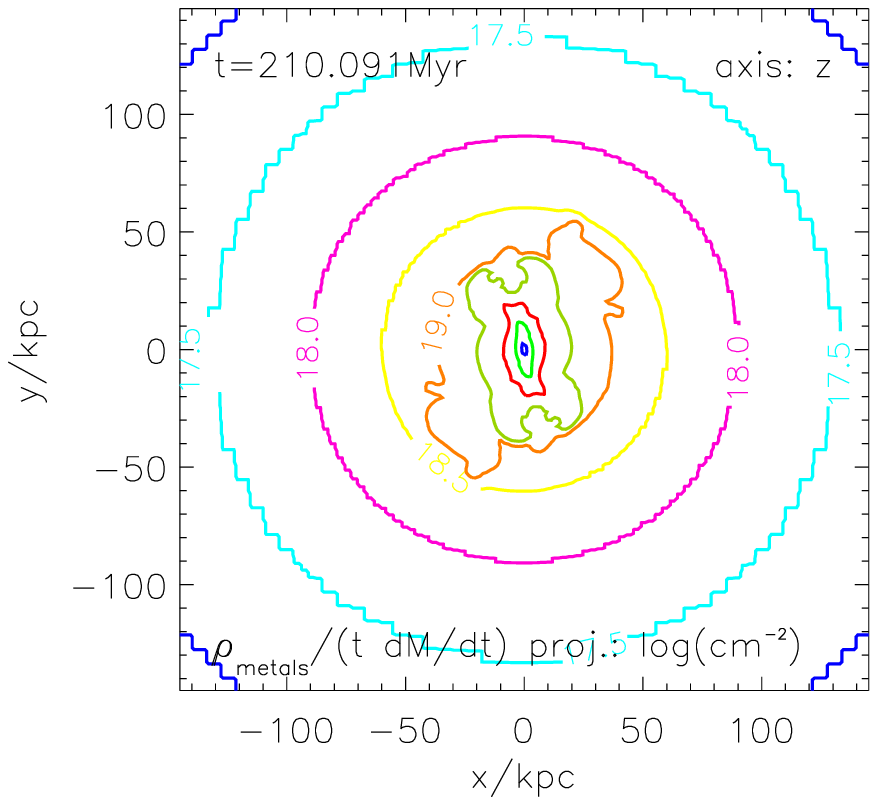}\vspace*{-1cm}\newline
\includegraphics[width=0.32\textwidth]{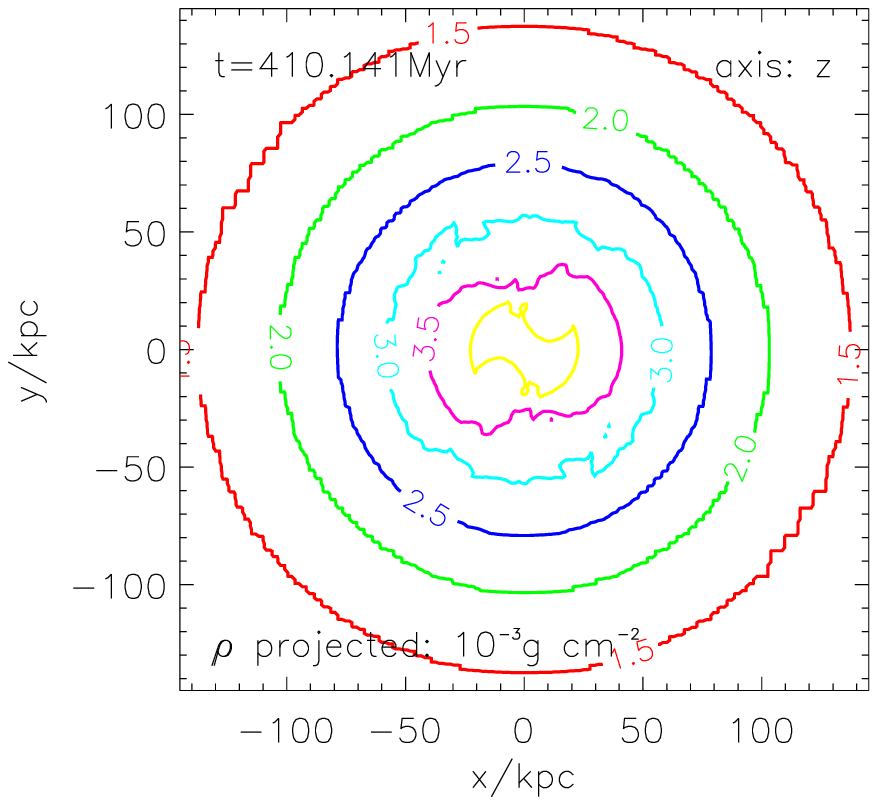}
\includegraphics[width=0.32\textwidth]{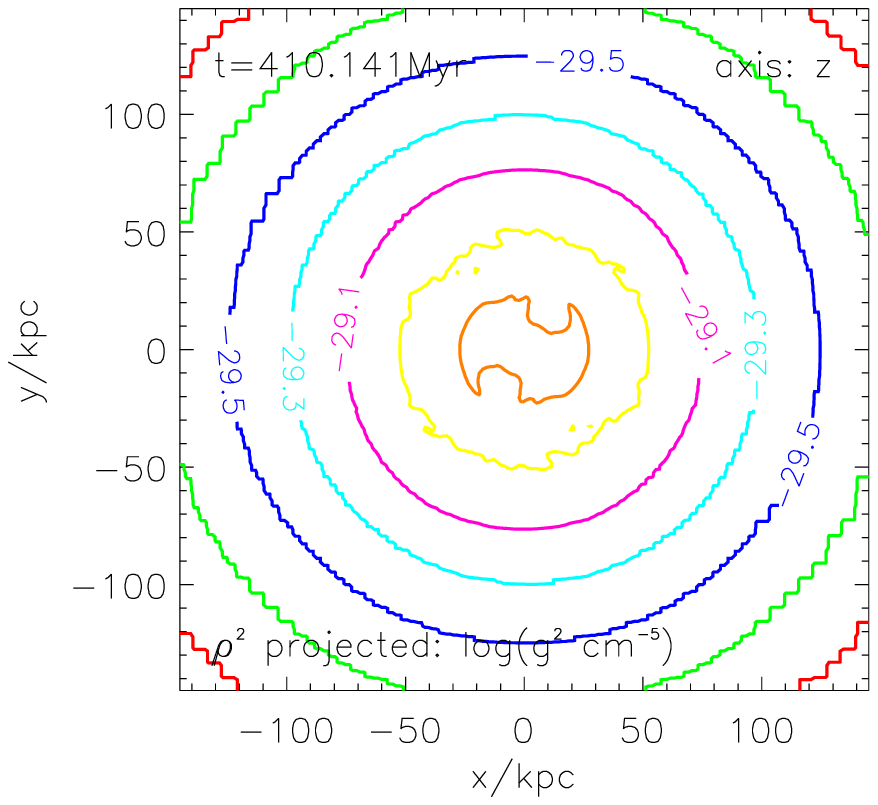}
\includegraphics[width=0.32\textwidth]{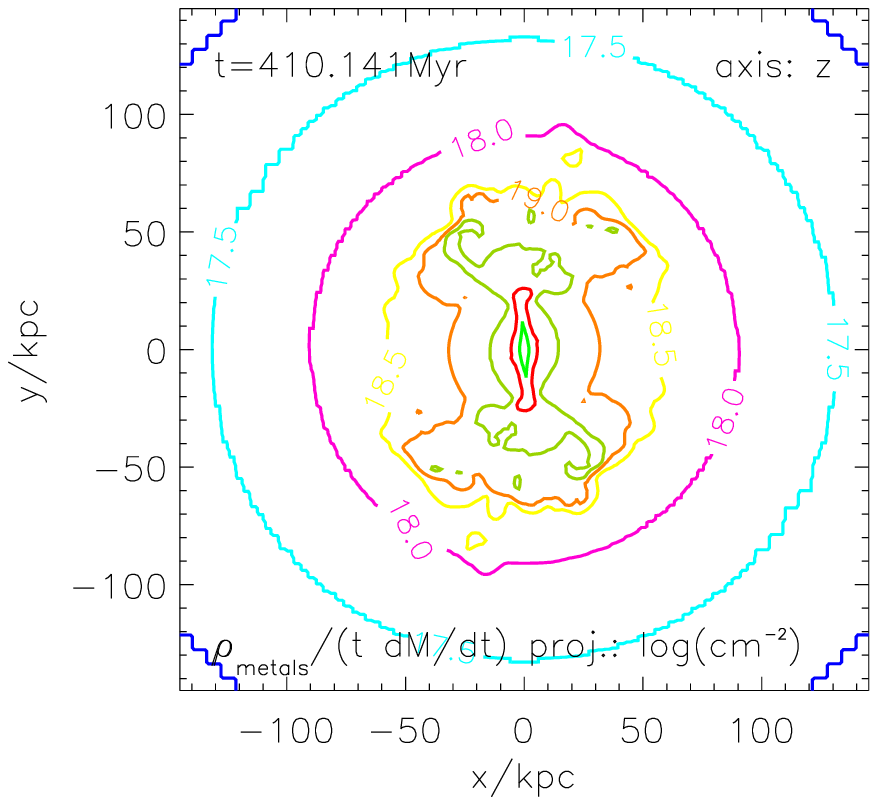}\vspace*{-1cm}\newline
\includegraphics[width=0.32\textwidth]{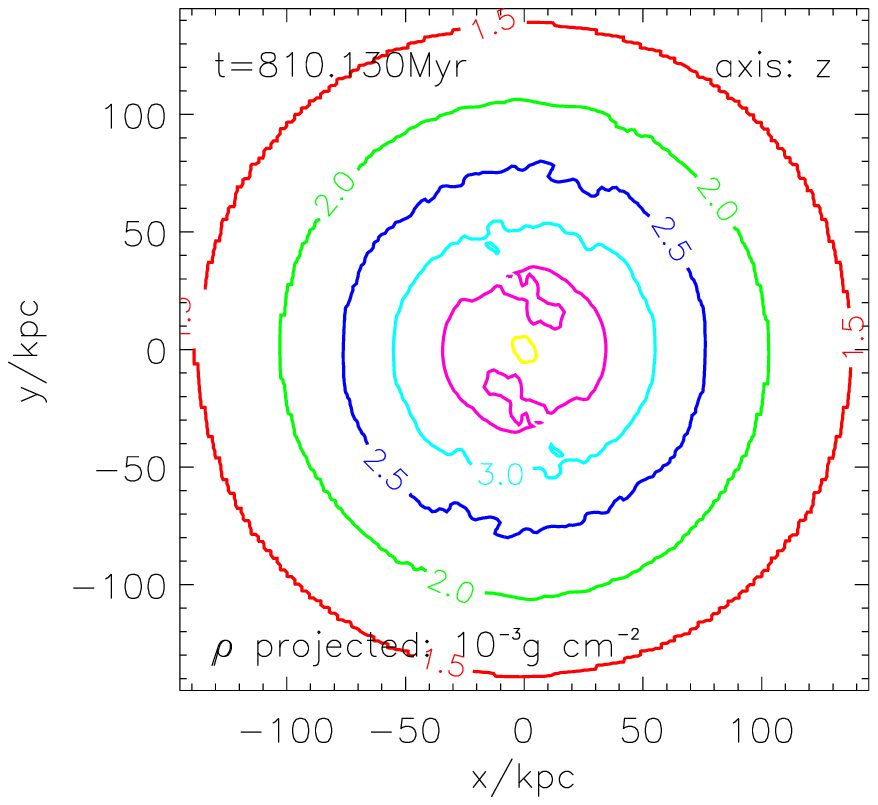}
\includegraphics[width=0.32\textwidth]{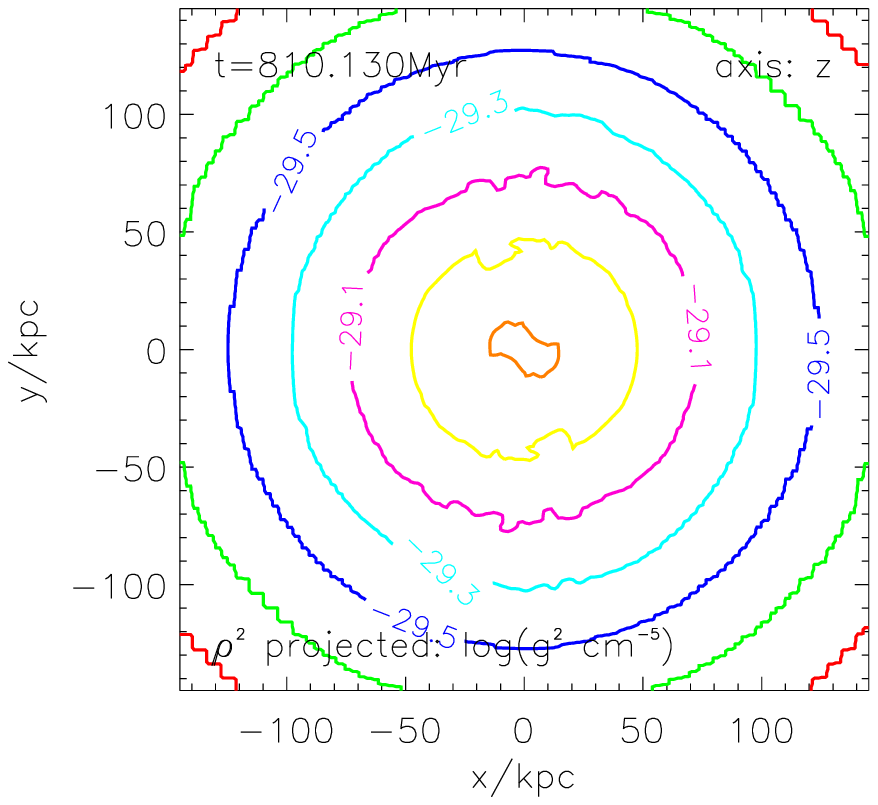}
\includegraphics[width=0.32\textwidth]{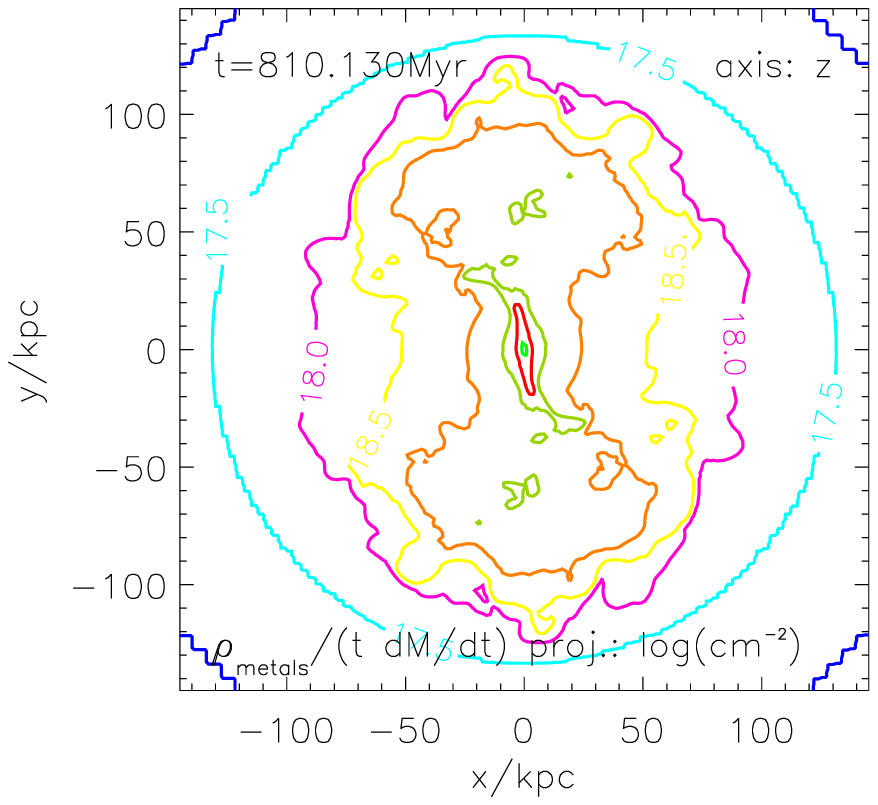}
\caption{Projections of gas density (left column), $\rho^2$ (middle
  column) and normalised metal density (right column) for
  different timesteps (see label in each panel) for simulation EVAC. Projection is done along the $z$-axis.}
\label{fig:proj_evac}
\end{figure*}

In Fig.~\ref{fig:prof_Mmets_evac_infl}, we investigate the sensitivity
of the metal profiles to the way we produce the bubbles.  We generate
the bubbles by either injecting hot gas into a number of computational
cells or by removing gas mass from a spherical region of space while
keeping the pressure constant.  Reassuringly, the differences between
the two methods are small. In the evacuation method the bubbles seem
to be able to carry metals out a little bit further. Also the global
effect (flattening of cumulative metal mass profile, see
Fig.~\ref{fig:prof_Mmets_evac_infl}) is a bit stronger in the
evacuation scenario. However, all properties concerning the transport
of metals in the cluster are very similar in both runs. The evacuation
method is computationally slightly cheaper because the timesteps
during the inflation phase become quite small. Projections of the
gas density, the X-ray surface brightness and the normalised metal
density for different timesteps are shown in
Fig.~\ref{fig:proj_evac}. Therefore, we produce the bubble by
evacuation in all our other runs.


\subsubsection{Resolution test}

\begin{figure*}
\vspace*{-0.5cm}
\includegraphics[width=0.4\textwidth]{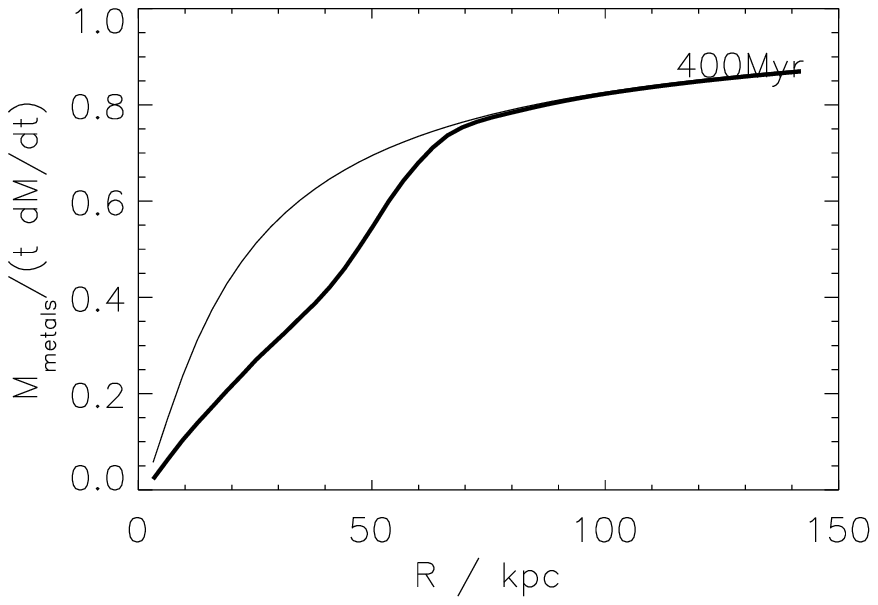}
\includegraphics[width=0.4\textwidth]{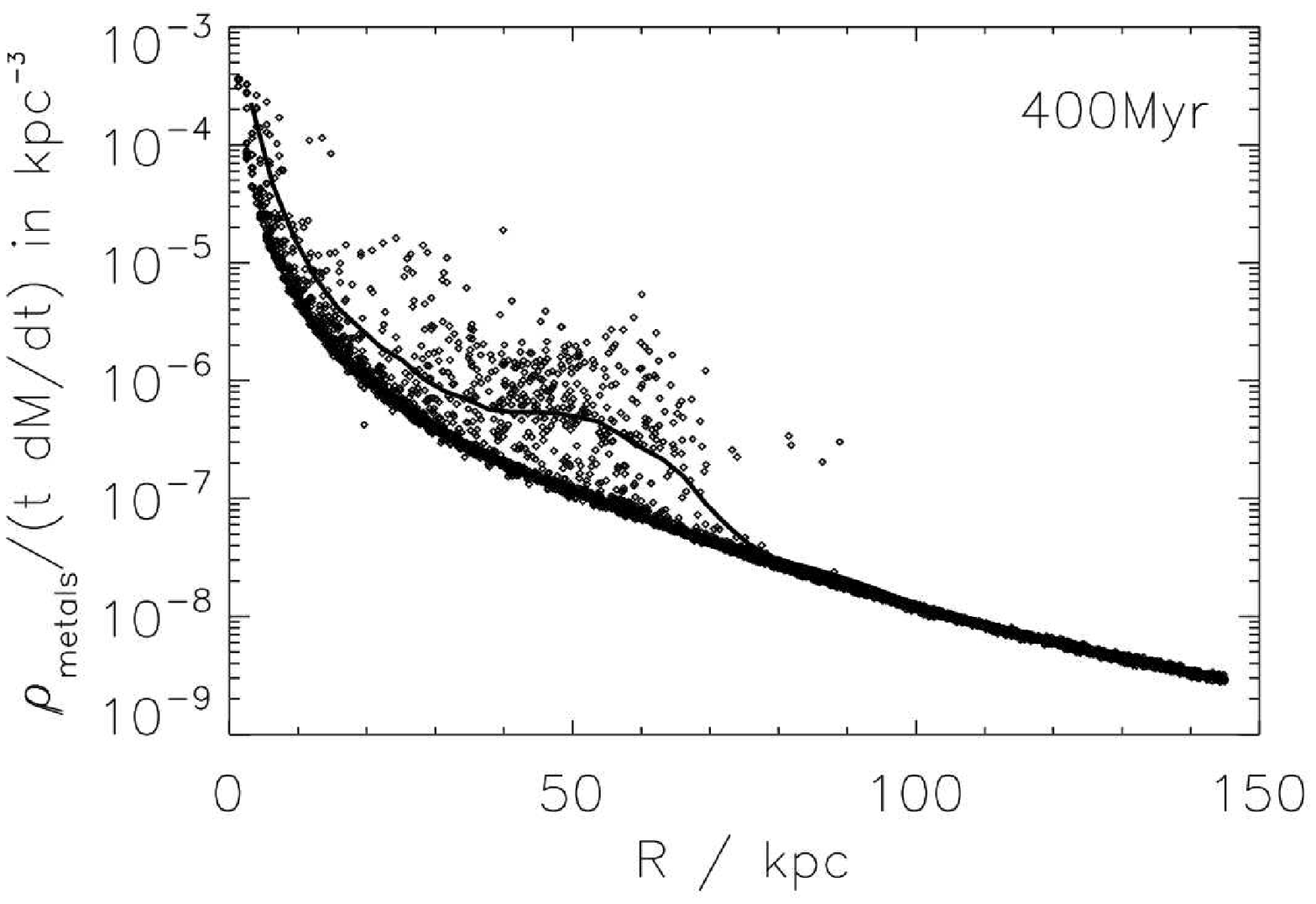}\\
\includegraphics[width=0.4\textwidth]{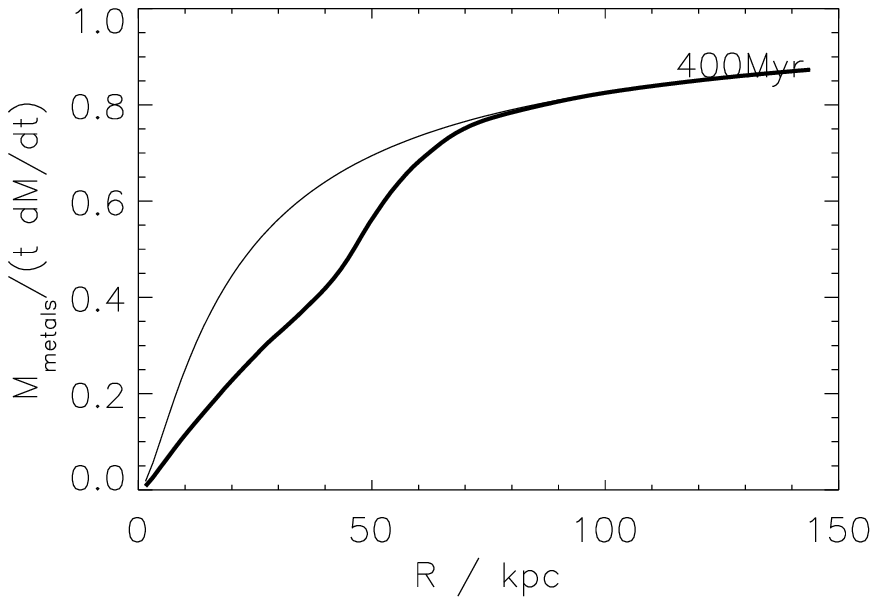}
\includegraphics[width=0.4\textwidth]{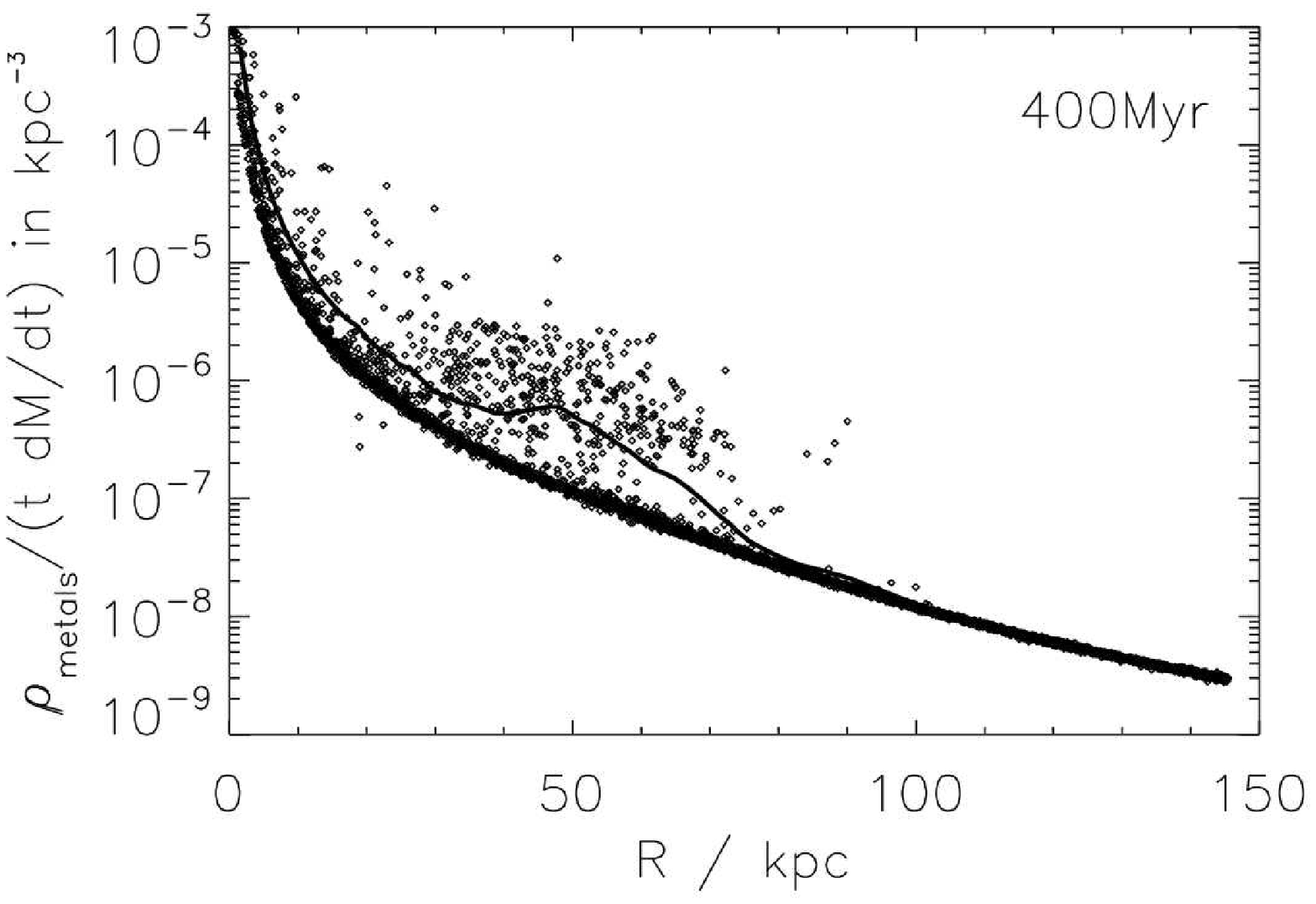}

\caption{Left: Normalised cumulative metal mass profile (metal mass inside
$r$) for run EVAC with an effective resolution of $192^3$ zones (top)
and for run EVAC\_HR with an effective resolution of $384^3$ zones
(bottom). The upper line in both plots corresponds to the case without
mixing. Right: Radial metal density profile normalised
to $t\dot M\Metal{}_0$ for the $192^3$ run (top) and the $384^3$ run
(bottom). All data corresponds to a time of 400 Myrs after the start
of the AGN activity.}
\label{fig:convergence}
\end{figure*}

In these kinds of simulations, it is essential to assess the errors
caused by the spatial discretisation. We can compare runs EVAC and
EVAC\_HR which have an effective resolution of $192^3$ and $384^3$
zones respectively. The morphologies of the bubbles are very similar
up to $250\Myr$ and the cumulative mass profiles are practically
indistinguishable. In Fig.~\ref{fig:convergence}, we show the
normalised cumulative metal mass profiles for runs with resolutions of
$192^3$ zones and $384^3$ zones. The resultant metal profiles suggest
that the simulation have converged numerically.

\subsection{The case of Perseus}

Having discussed a somewhat generic cluster model above, we now turn
to the brightest X-ray cluster A426 (Perseus) that has been studied
extensively with {\sc Chandra} and {\sc XMM}-Newton. Our simulations
of the Perseus cluster show two things: (i) In
the case of Perseus, the metal distribution is somewhat broader. This
is a result of the greater pressure gradient in Perseus which leads to
a greater buoyancy force and thus a higher velocity of the
bubbles. (ii) We find that despite varying the initial positions of
the bubbles, the resulting metal distribution remains fairly elongated
and decidedly non-spherical. 

\begin{figure*}
\includegraphics[trim=0 80 0 0,clip,width=0.45\textwidth]{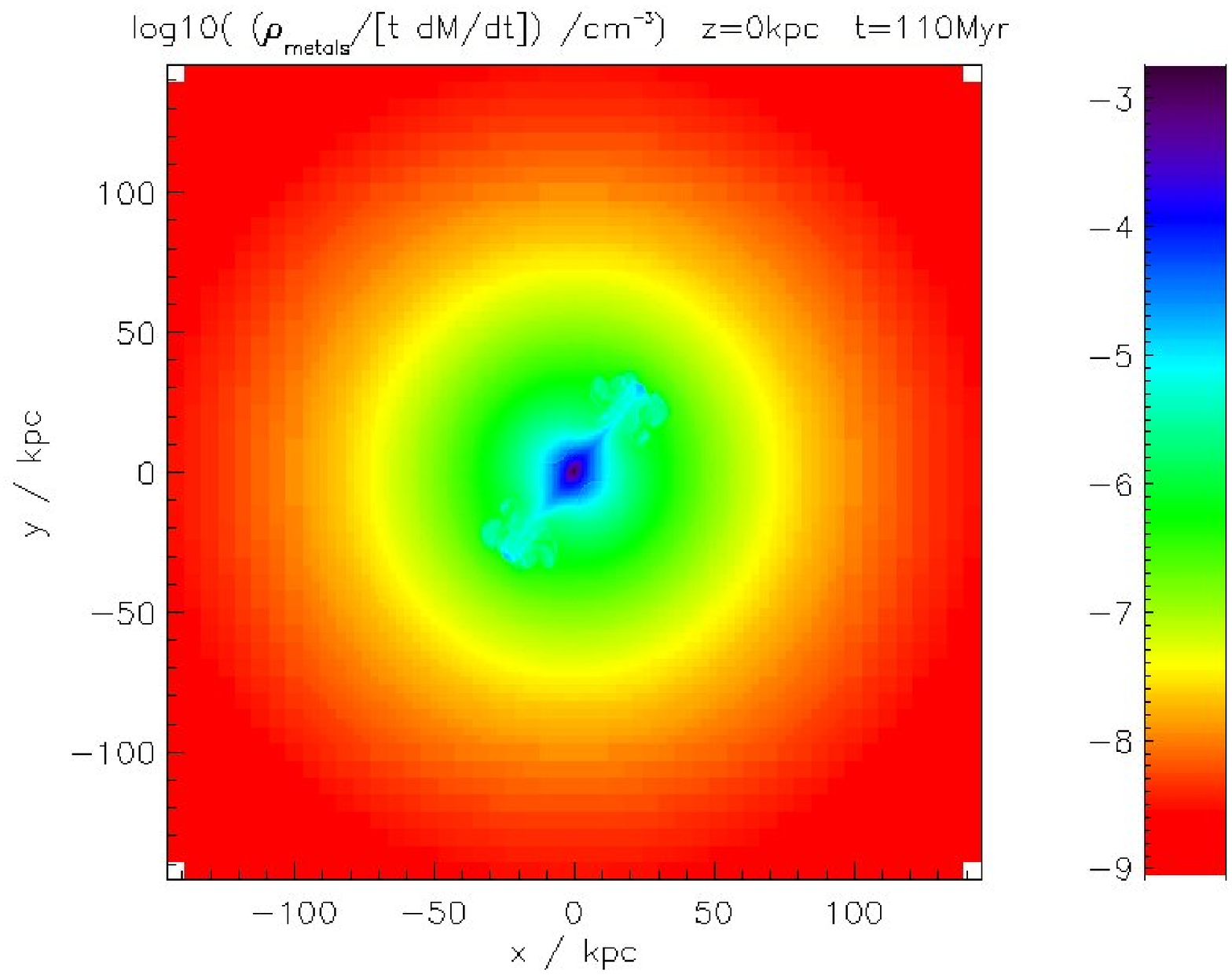}
\includegraphics[trim=0 80 0 0,clip,width=0.45\textwidth]{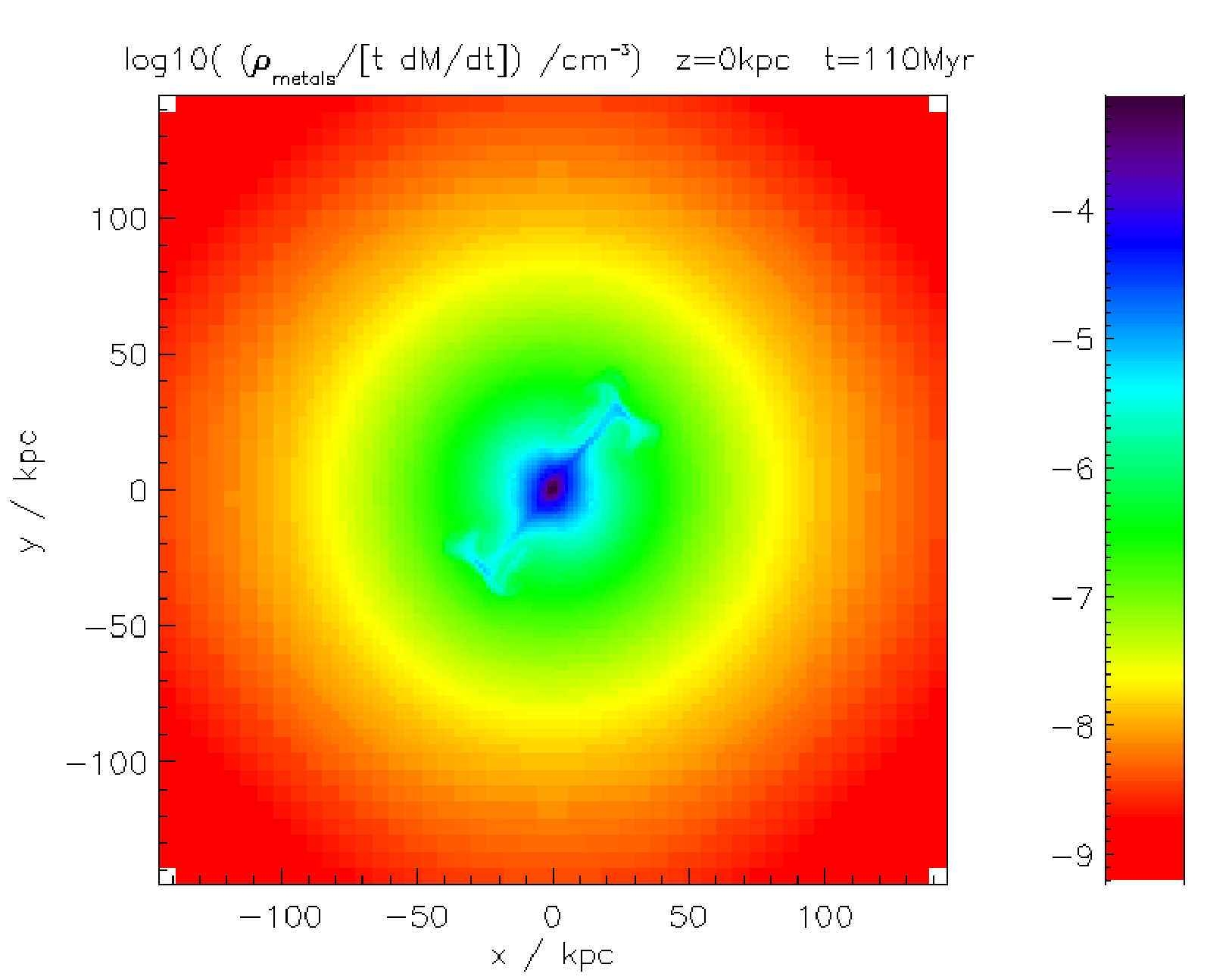}\newline
\includegraphics[trim=0 80 0 0,clip,width=0.45\textwidth]{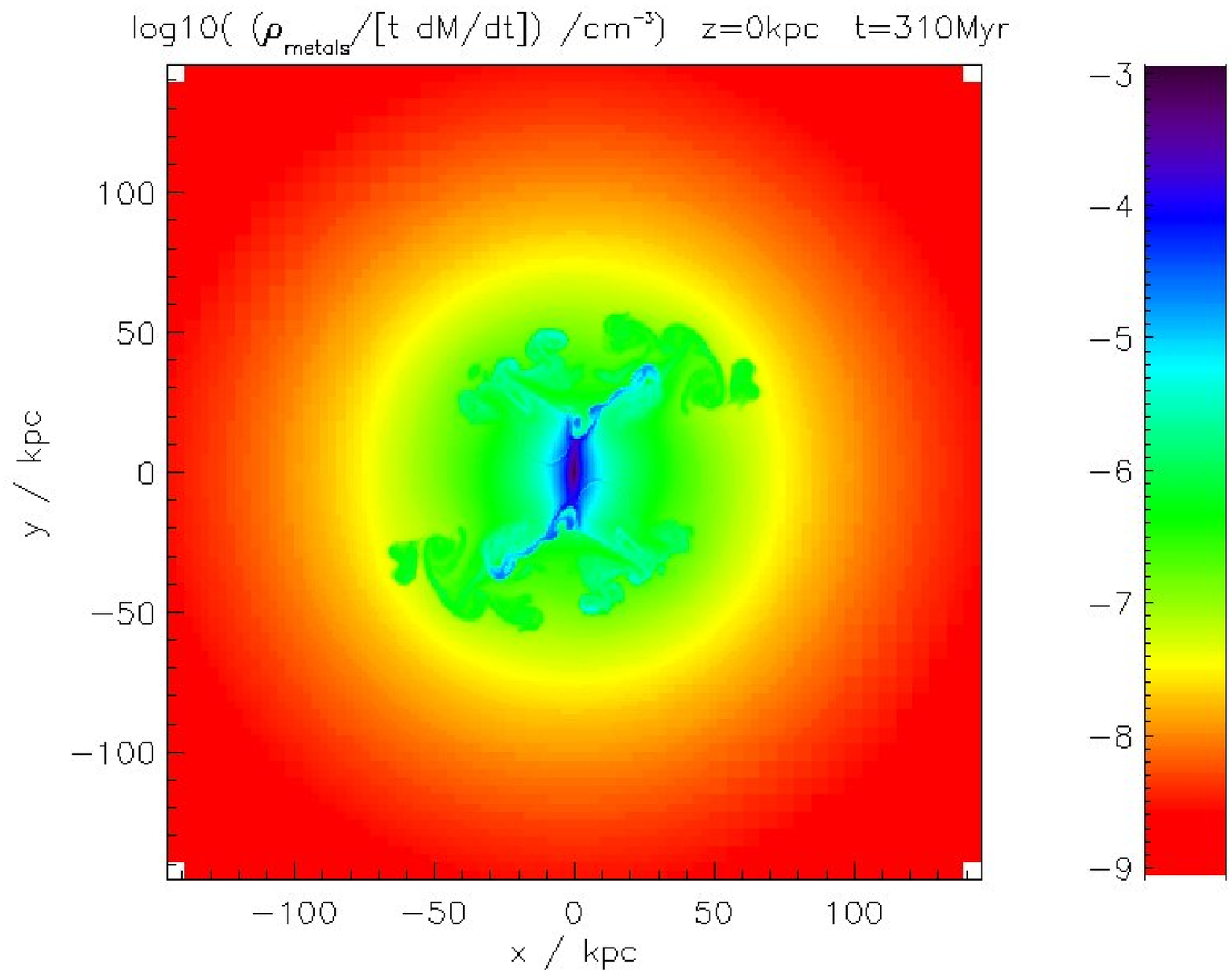}
\includegraphics[trim=0 80 0
0,clip,width=0.45\textwidth]{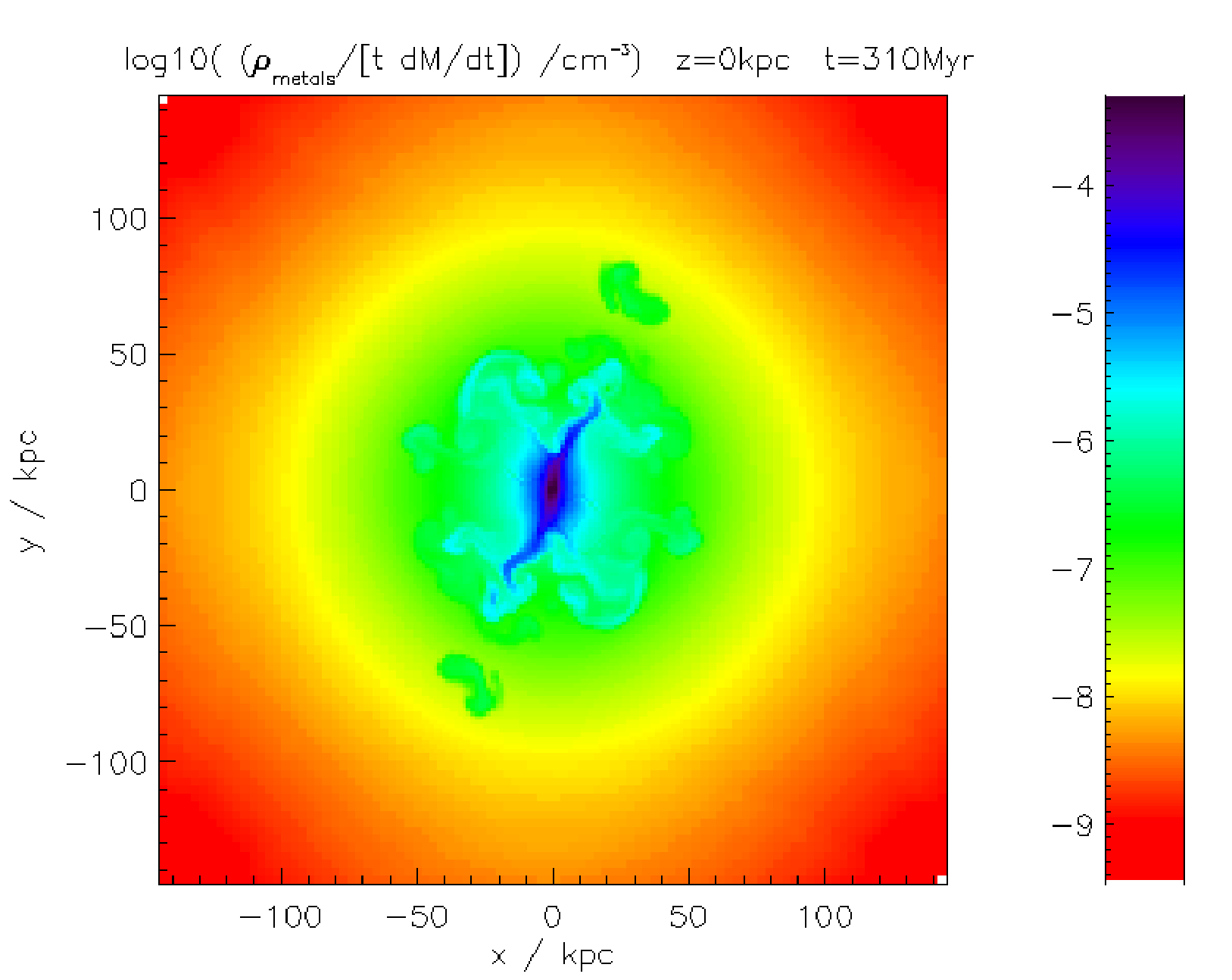}
\includegraphics[trim=0 80 0 0,clip,width=0.45\textwidth]{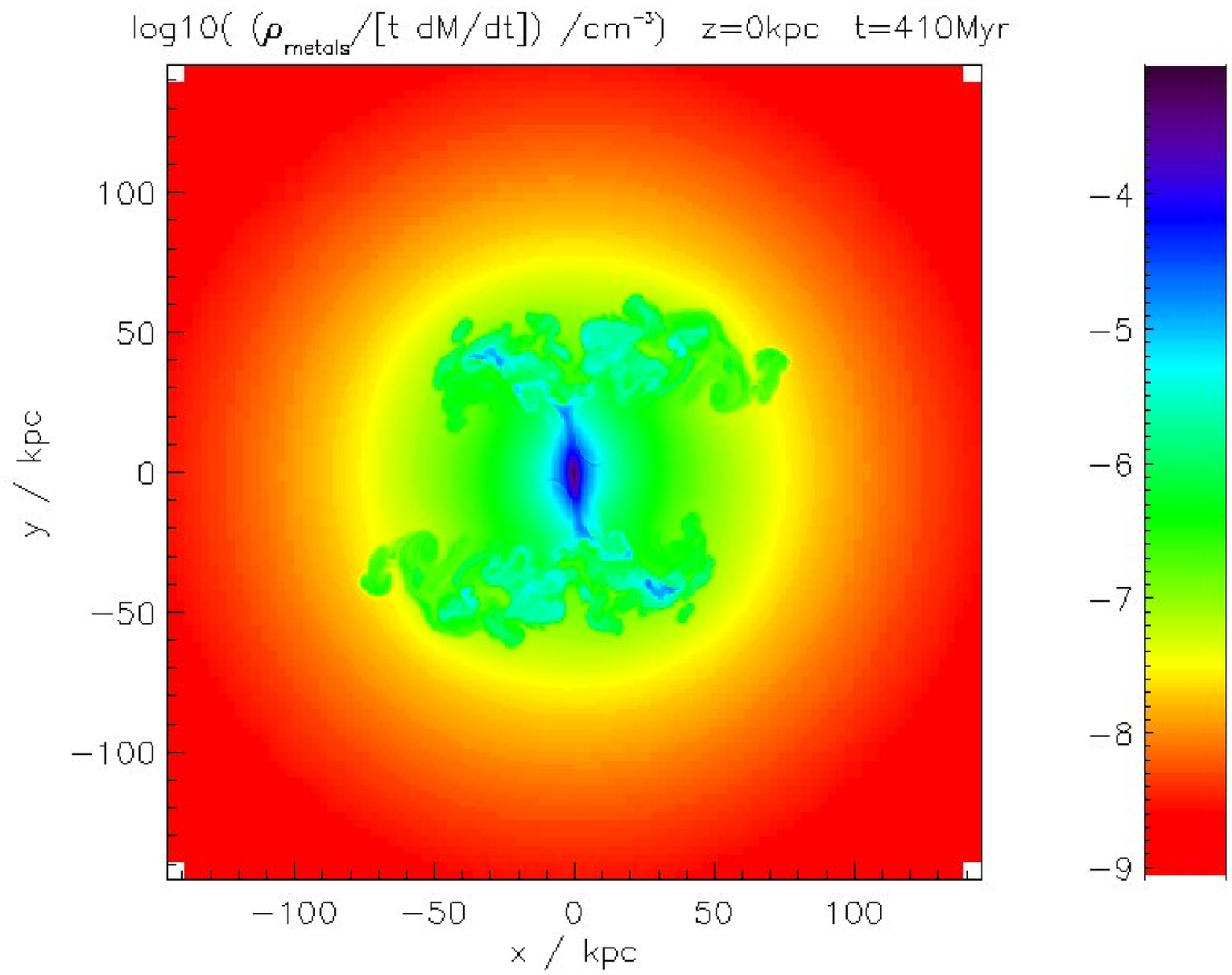}
\includegraphics[trim=0 80 0 0,clip,width=0.45\textwidth]{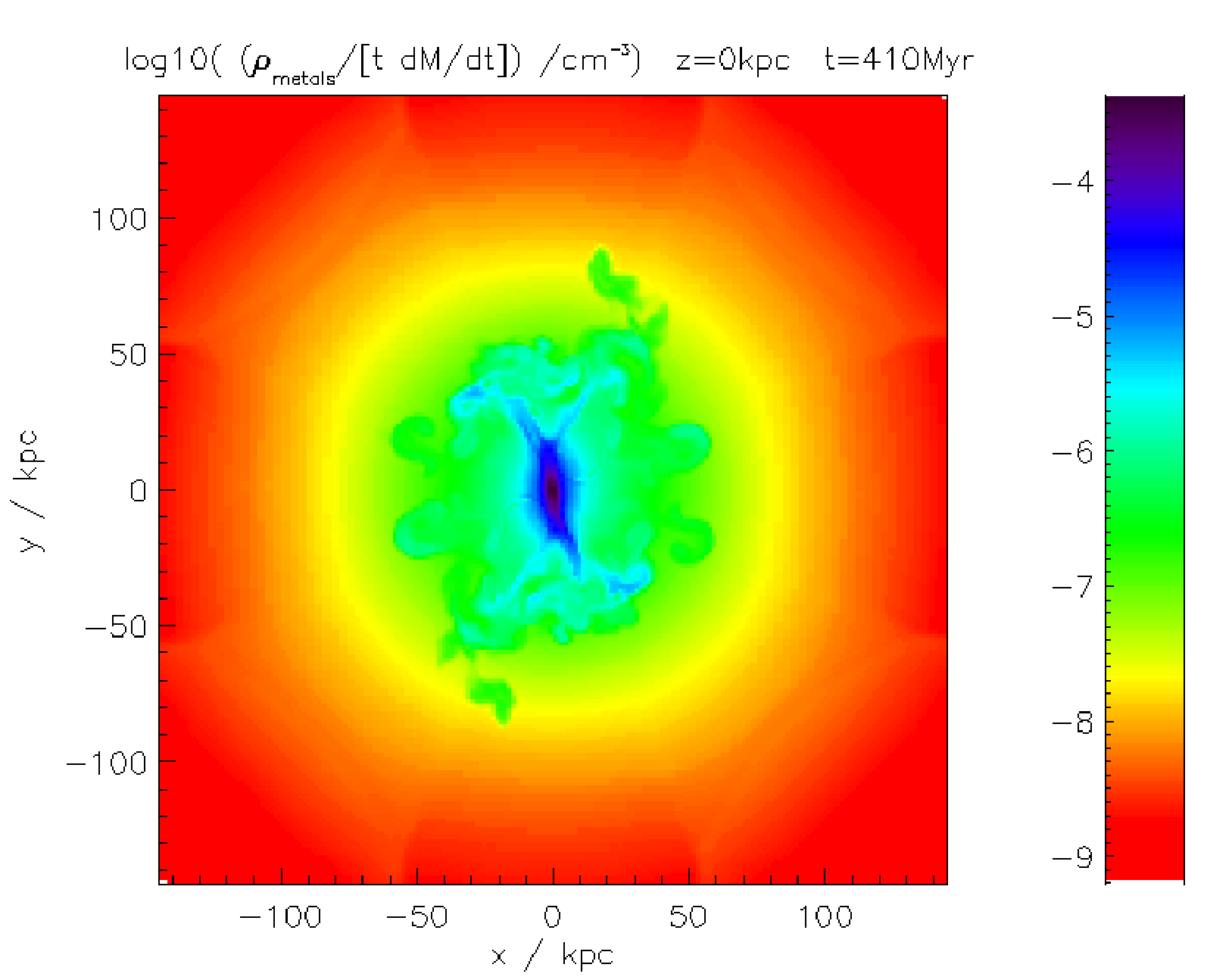}
\caption{Slice through the computational box showing local metal
fraction at three different times. Left
  column is for the generic background cluster (EVAC), and right
  column for the simulation based on Perseus (PERSEUS\_50).}
\label{fig:slice_mfrac_evac_pers}
\end{figure*}

Fig.~\ref{fig:slice_mfrac_evac_pers} shows the local metal fraction at
three different times in slices through the centre of the cluster. The
left column is for the generic background cluster (EVAC) and right
column for the simulation based on Perseus. For the Perseus model, the
metal disribution is somewhat broader than in our generic
model. However, all qualitative features discussed above for the
generic model remain valid for the cluster based on Perseus.




%% file: diff_const.tex
\subsection{Estimate of diffusion constant} \label{sec:diff_const}
In this section, we wish to parametrise the transport of metals by computing
an effective diffusion constant. We will begin by defining a cumulative metal
mass profile
%
\begin{equation}
M\Metal(R)=\int\limits_{r<R} \rho\Metal(\vec r) \mathrm{d}V 
\end{equation}
%
and the averaged metal density profile
%
\begin{equation}
\bar\rho\Metal(r)= \frac{1}{4\pi r^2} \frac{\partial M\Metal(r)}{\partial r} .
\end{equation}
%
On the basis of these quantities, we can estimate a diffusion constant
that can be compared with \cite{rebusco:05}. We start from the
spherically symmetrical diffusion equation including the metal source
term
%
\begin{equation}
\frac{\partial \bar\rho\Metal(r,t)}{\partial t} = \frac{\partial}{\partial r}
D \frac{\partial}{\partial r}\bar\rho\Metal(r,t) + \dot\rho\Metal(r) ,
\end{equation}
%
where $D$ is the diffusion constant, and $\dot\rho\Metal(r)$ is the metal
source profile according to Eq.~\ref{eq:hernquist}. 
Discretising this between the radii $R_1$ and $R_2$ (i.e. $\Delta R=R_2-R_1$ and
$\bar R= (R_1 + R_2)/2$) and timesteps $t_1$ and $t_2$ ($\Delta t=t_2-t_1$ and
$\bar t= (t_1 + t_2)/2$) yields
%
\begin{eqnarray}
&&\frac{\bar\rho\Metal(\bar R,t_2) - \bar\rho\Metal(\bar R,t_1)}{\Delta t}
= \dot\rho\Metal(\bar R) + \nonumber\\
&&D(\bar R, \bar t)\;\frac{\frac{\partial}{\partial r}\bar\rho\Metal(R_2,\bar t) - \frac{\partial}{\partial r}\bar\rho\Metal(R_1,\bar t)}{\Delta R},
\end{eqnarray}
where we have assumed that $D$ varies only slowly with radius,
i.e. $D(R_1,t)\approx D(R_2,t) \approx D(\bar R, t)$. Reordering yields
%
\begin{eqnarray}
&& D(\bar R, \bar t) = \nonumber\\
&&\Delta R \;\frac{\bar\rho\Metal(\bar R,t_2) -
  \bar\rho\Metal(\bar R,t_1) - \dot\rho\Metal(\bar R)\,\Delta t}{\Delta
  t\;\left[\frac{\partial}{\partial r}\bar\rho\Metal(R_2,\bar t) - \frac{\partial}{\partial r}\bar\rho\Metal(R_1,\bar t)  \right]}
\end{eqnarray}

\begin{figure*}
\includegraphics[width=0.45\textwidth]{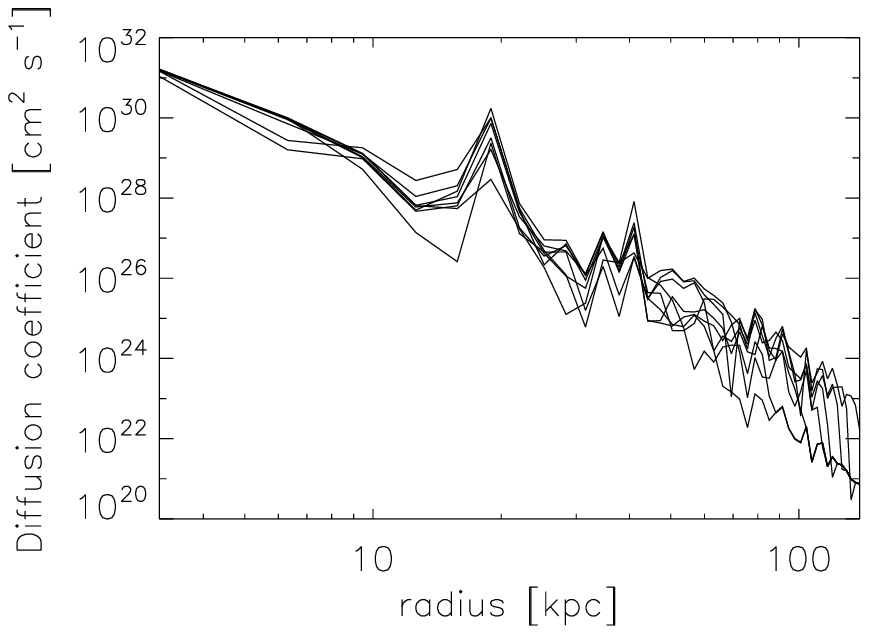}
\includegraphics[width=0.45\textwidth]{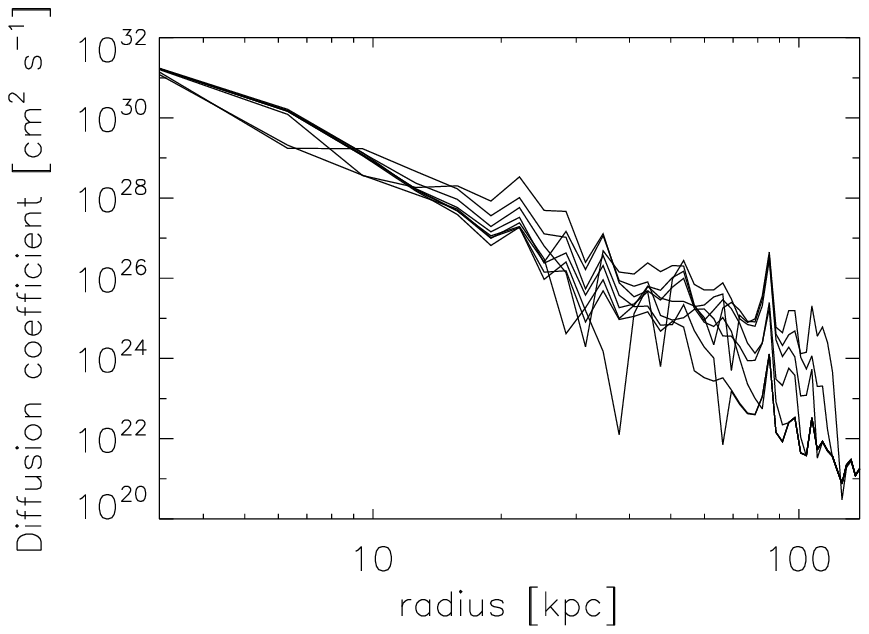}\\
\includegraphics[width=0.45\textwidth]{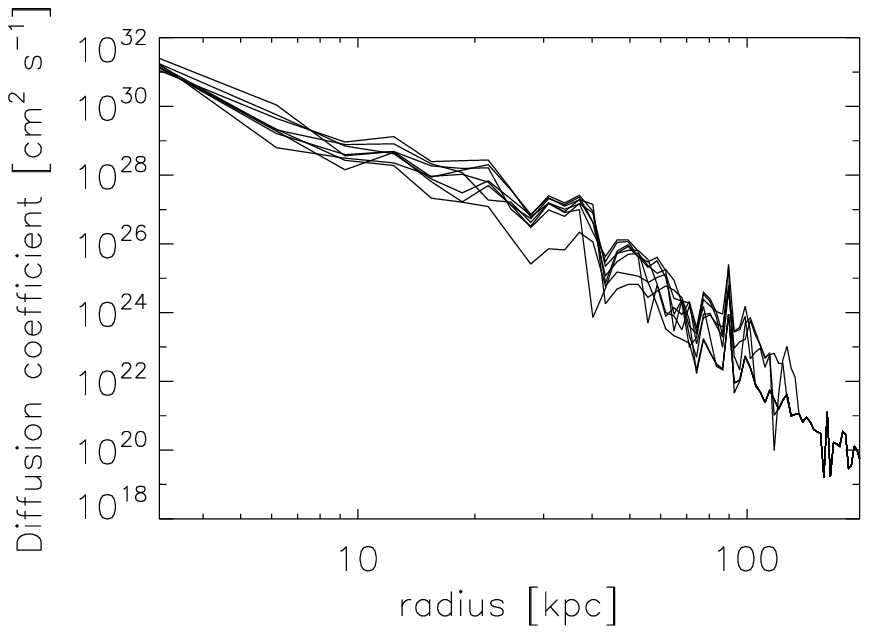}
\includegraphics[width=0.45\textwidth]{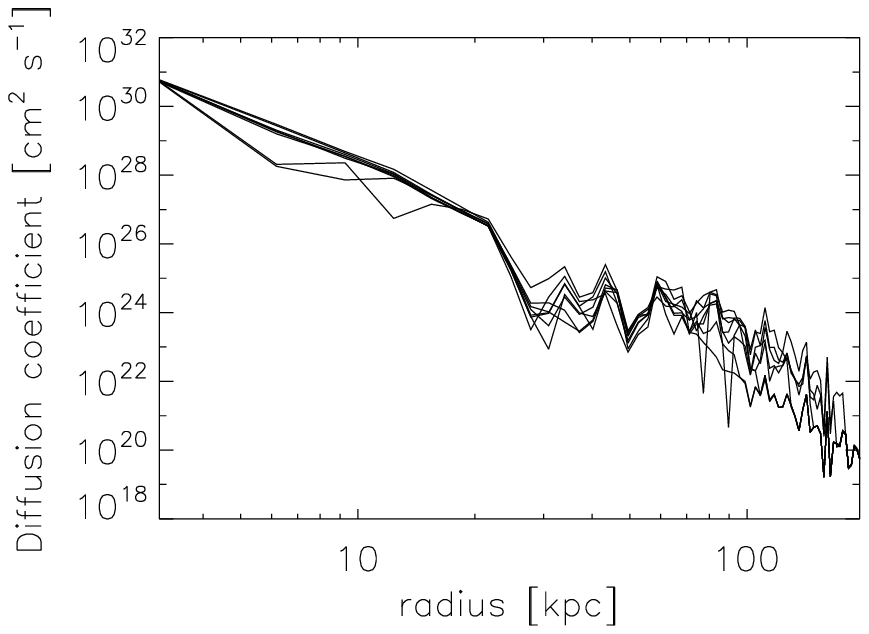}\\
\includegraphics[width=0.45\textwidth]{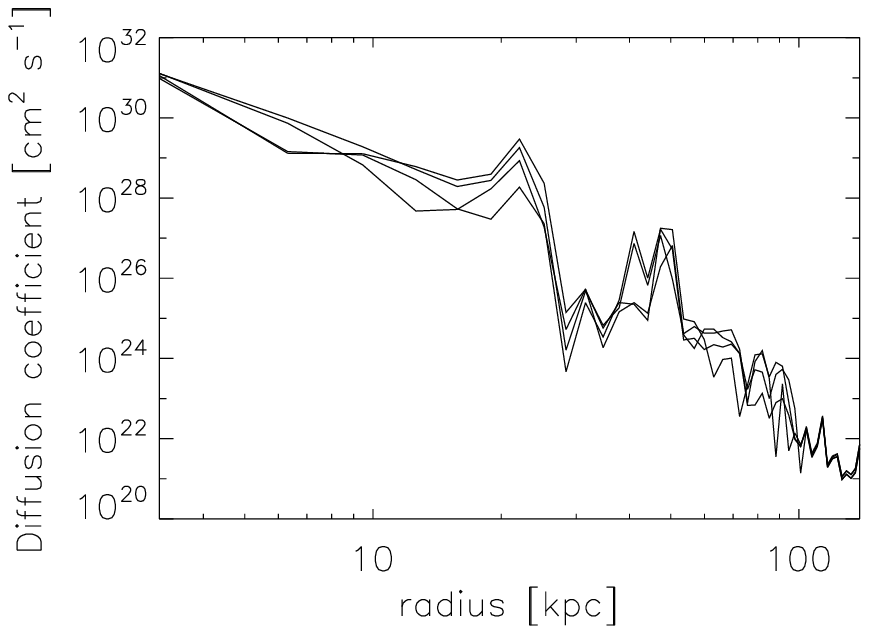}
\includegraphics[width=0.45\textwidth]{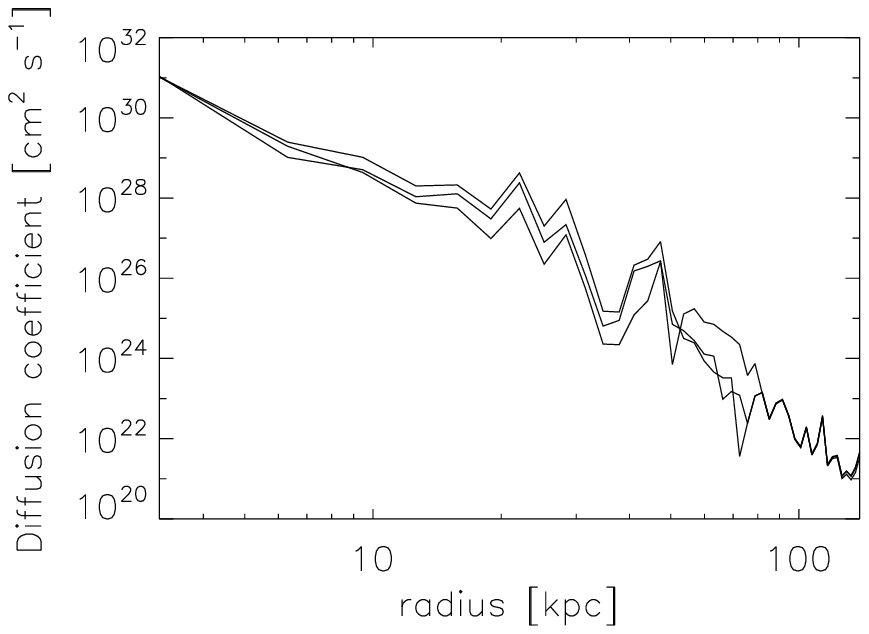}
\caption{Diffusion coefficients for the runs EVAC (top left), INFL (top
  right), TAUBBL200, (middle left), DISTCTR (middle right),
  PERSEUS\_50 (bottom left) and PERSEUS\_200 (bottom right). The different
  curves denote different times.}
\label{fig:diffconst}
\end{figure*}

For the runs described in the previous sections, we have computed the radial
profiles of the diffusion constants in Fig.~\ref{fig:diffconst}.

%% file: discussion.tex
\section{Discussion}
%
In our simulations, the dynamics of the ICM is entirely driven by
underdense bubbles that are initially at rest and then start to rise
by the action of buoyancy. This is clearly a simplified picture. In
reality, these bubble are inflated by relativistic jets whose physics
is still very poorly understood. Real AGN jets have a dynamics that
can only be fully captured with special-relativistic MHD codes that
need to resolve scales much below the scales resolved in our
simulations.

However, bubbles such as those set up in our simulations are similar
to the bubbles observed in galaxy clusters such as the Perseus
cluster. The bubbles in Perseus and other clusters are nearly in
pressure equilibrium with the ambient medium, they are nearly
spherical and move subsonically through the ICM. Thus, we ignore the
entire process that has led to the formation of these bubbles and
focus solely on the buoyancy-driven motions of the bubbles.\\

There have been some attempts to simulate the inflation
of the bubbles by a jet and assess its consequences for the ICM.
\cite{omma:04} simulate the effect of slow jets on a cool core in a galaxy cluster.
Simulations by \cite{vernaleo:06} show that jets launched in a
hydrostatic cluster fail to heat the cluster on long time scales
because the jet forms a low-density channel through which the material
can flow freely, carrying its energy out of the cooling core.\\

If bubbles in a hydrostatic cluster are launched repeatedly at the
same position, the resulting metal distribution becomes very
aspherical and elongated along the axis of the bubble. If, however the
bubbles are launched at changing positions, the metal distribution is
somewhat isotropised. However, as is apparent e.g. from
Fig.~\ref{fig:slice_mfrac_evac_pers}, even when we vary the bubble
position within a cone of $\sim 30\deg$, the resulting metal
distribution remains elongated and non-spherical. Thus, if the metal
distribution should turn out to be spherical, this would necessitate
additional motions caused for example by mergers or by stirring by
galaxies in order to isotropise the metals.

In real clusters, the ICM is not in hydrostatic equilibrium and can
show strong motions. Galaxies that fly through the cluster produce
motions and turbulence. Merger activity can stir the ICM even more
violently. These motions will all contribute to dispersing the metals
throughout the ICM. Moreover, density anisotropies can cause the
bubbles to move away from the axis of the AGN jet, as the bubbles
follow the local density gradient. Indeed, in quite a few clusters,
most famously in M87, the bubbles are widely dispersed through the
cluster and do not trace the alleged axis of the AGN
jet. \cite{heinz:06} have performed three-dimensional adaptive-mesh
simulations of jets situated at the centre of a cluster drawn from a
cosmological simulation. These simulations show that cluster
inhomogeneities and large-scale flows have a significant impact on the
morphology of the bubbles. Moreover, these motions help to prevent the
formation of channels along the bubbles axes as found by
\cite{vernaleo:06}.\\

The diffusion coefficients inferred from our simple experiments lie at
values of around $\sim 10^{29}$ cm$^2$s$^{-1}$ at a radius of 10
kpc. Right at the centre, they rise to about $\sim 10^{31}$
cm$^2$s$^{-1}$ and fall to about $\sim 10^{25}$ cm$^2$s$^{-1}$ at a
radius of 100 kpc. Modulo a factor of a few, the coefficients for all
runs lie around these values. Differences are induced by the bubble
sizes, the initial positions, the recurrence times and the pressure
profile in the cluster. Interestingly, the runs modelled on the
Perseus cluster yield diffusion coefficients that agree very well with
those estimated by assuming a simple diffusion model (see
\cite{rebusco:05}). As we do not take into account motions that may
have been induced by mergers etc., the values for the diffusion
coefficients constitute lower limits on the real values. Thus, we can
conclude that AGN-induced motions are sufficient to explain the broad
abundance peaks in clusters. It is also striking that the diffusion
coefficients fall so steeply with radial distance from the
centre. Over two decades in radius they decrease by almost 10 orders
of magnitude. Very approximately, they fall like $r^{-5}$. This strong
radial dependence of the transport efficiency is, partially, a result
of the three-dimensional nature of the bubble-induced motions,
i.e. uplifted metals are diluted over the entire radial shell.  A
second effect is the decaying ``lift'' power of the bubbles as they
rise. The bubbles accelerate only at the beginning of their ascent
through the cluster. However, very quickly instabilities set in that
slow down the bubble until it finally comes to a halt at about 100 kpc
from the centre. The maximal lift power is attained when the bubbles
are still intact and have their largest velocity. This occurs at radii
of about 20 kpc - 30 kpc, and this coincides with peaks in the
diffusion coefficient as can be seen from Fig.~\ref{fig:diffconst},
particularly for runs EVAC and PERSEUS\_50.

Finally, we may add that if hydrodynamic instabilities are suppressed
by magnetic fields or viscous forces, for example, the bubbles would
stay intact for longer and could rise to greater distances thus also
enhancing the diffusion coefficients at larger radii.

%% file: summary.tex
\section{Summary}
%

In a series of numerical experiments, we have performed a systematic
study on the effect of bubble-induced motions on metallicity profiles
in clusters of galaxies. In particular, we have studied the dependence
on the bubble size and positions, the recurrence times of the bubbles,
the way these bubbles are inflated and the underlying cluster profile.

On the basis of our 3D hydro simulations, we wish to make the
following main points:

\bi

\ii Larger bubbles lead to mixing to larger radii. For realistic
parameters and in the absence of stabilising forces, bubbles hardly
move beyond distances of 150 kpc from the centre.

\ii Mixing scales roughly with bubble frequency.

\ii The metal distribution is not sensitive to the way of how bubbles are inflated.

\ii In hydrostatic cluster models, the resulting metal distribution is
very elongated along the direction of the bubbles. Anisotropies in the
cluster and/or ambient motions are needed if the metal distribution is
to be spherical.

\ii The diffusion coefficients inferred from our simple experiments
lie at values of around $\sim 10^{29}$ cm$^2$s$^{-1}$ at a radius of 10
kpc.

\ii The runs modelled on the Perseus cluster yield
diffusion coefficients that agree very well with those inferred from
observations (see \cite{rebusco:05}).

\ei